\documentclass[journal]{IEEEtran}
\usepackage{amsfonts}
\usepackage{makecell}
\usepackage{diagbox} 
\usepackage{boldline}
\usepackage{multicol}
\usepackage{multirow}
\usepackage{subfiles}
\usepackage{xcolor}

\usepackage{cite}

\ifCLASSINFOpdf
  \usepackage[pdftex]{graphicx}
\else
\fi
\usepackage[cmex10]{amsmath}

\usepackage{stfloats}
\usepackage{url}

\usepackage{orcidlink}
\usepackage{hyperref}

\usepackage[switch,pagewise]{lineno}

\begin{document}
\title{Calibration of the Polarimetric GNSS-R Sensor in the Rongowai Mission}
%
%
%

\author{Dinan~Bai \orcidlink{0009-0009-6500-4372},~\IEEEmembership{Member,~IEEE,}
        Christopher S. Ruf \orcidlink{0000-0002-5937-4483},~\IEEEmembership{Life Fellow,~IEEE,}
        Delwyn Moller \orcidlink{0000-0003-4207-1539},~\IEEEmembership{Senior Member,~IEEE}
\thanks{D. Bai and C. Ruf are with the Department
of Climate and Space Science and Engineering, University of Michigan, Ann Arbor,
MI, 48105 USA e-mail: dinanbai@umich.edu}
\thanks{D. Moller is with the Department of Electrical and Computer Engineering, University of Auckland, Auckland, New Zealand.}
}

%
%

\markboth{Transaction on Geoscience and Remote Sensing, 2025}%
{Shell \MakeLowercase{\textit{et al.}}: Bare Demo of IEEEtran.cls for Journals}
%



\maketitle

\begin{abstract}
Polarimetric GNSS-R systems, equipped with an additional polarization channel, offer enhanced capabilities for separating vegetation and surface scattering effects, thereby improving GNSS-R land remote sensing applications such as soil moisture retrieval in vegetated and forested areas and biomass estimation. However, the effectiveness of these applications relies on accurate calibration of the polarimetric GNSS-R sensor. In the Rongowai mission, a newly developed Next Generation GNSS-R Receiver (NGRx) is installed on a domestic Air New Zealand airplane to collect data during its commercial flights. The NGRx processes multi-GNSS satellite signals simultaneously and utilizes a dual-channel (LHCP and RHCP) antenna, thereby improving spatial coverage and retrieval accuracy. The dual-polarized antenna also provides the possibility to examine the polarimetric GNSS-R system. In this article, a new methodology is developed to calibrate the Level-1 power measurement and the on-board antenna cross-pol gain by comparing measurements from inland lakes and ocean with modeled results. The calibration results in a 34\% decrease in the uncertainty in co-pol reflectivity retrieval. The retrieved cross-pol and co-pol reflectivity after calibration are examined by their statistical distribution and spatial mapping with 1.5 km resolution, with multi- land surface types and incidence angles. These results validate the effectiveness of the calibration method and pave the way for future terrestrial science applications.
\end{abstract}

\begin{IEEEkeywords}
Remote Sensing, Polarimetric GNSS-R, Calibration, Rongowai.
\end{IEEEkeywords}

%
\IEEEpeerreviewmaketitle


\section{Introduction}
%
%
%
%
\IEEEPARstart{S}{oil}  moisture is a critical component in global hydrological and geophysical processes, influencing land-atmosphere interactions, climate patterns, and agricultural practices \cite{soil_moisture_1, soil_moisture_2, soil_moisture_3}. Consequently, the ability to remotely sense soil moisture on a global scale is of paramount importance. Microwave remote sensing techniques exploit their sensitivity to the change of soil dielectric constant at microwave frequencies due to the proportion of water mixed inside soil to retrieve soil moisture. Spaceborne remote sensing systems including passive radiometers (SMAP \cite{SMAP_mission}, SMOS \cite{SMOS_mission}) and active radars (Sentinel-1, ENVISAT/ASAR, and TerraSAR-x1 \cite{Setinal_1_SM, ENVISAT_SM, TerraSAR_SM}) have been developed for global soil moisture monitoring. In recent decades, a new microwave remote sensing system, Global Navigation Satellite System Reflectometry (GNSS-R) which measures opportunistic GNSS signals scattered from the earth surface has gained increasing interest. Spaceborne GNSS-R has been investigated in satellite missions such as UK-DMC \cite{DMC_mission} and TDS-1 \cite{TDS_mission} launched in 2004 and 2014 respectively. The NASA Cyclone Global Navigation Satellite System Earth Venture mission (CYGNSS) \cite{cygnss_mission} which consists of eight satellites was launched in 2016. CYGNSS was originally designed to measure surface windspeed in tropical cyclones to improve storm forecasting \cite{cygnss_wind}, but later extended to land remote sensing applications in 2019, following a series of studies \cite{GNSS_R_SM_sensitivity1,GNSS_R_SM_sensitivity2,  GNSS_R_SM_sensitivity3, GNSS_R_SM_sensitivity4} revealing the sensitivity of spaceborne GNSS-R to surface reflectivity measurement which is directly related to inland waterbody extent and near-surface soil moisture. Recent studies \cite{CYG_SM_Clara} have utilized CYGNSS observables for global soil moisture estimation with 6- to 24-hour time resolution, which makes it possible to bridge the gap for monitoring fast-evolving hydrological events such as drought/flood monitoring, wildfire forecast, and water and energy cycle monitoring. However, traditional GNSS-R such as CYGNSS have only one single Left-Hand Circular Polarization (LHCP) channel, designed to only measure the cross-pol (Right-Hand Circular Polarization (RHCP) incident, LHCP scattered) component of GNSS signals scattered from the Earth's surface. With only one single cross-pol channel the tangled effects of surface roughness, vegetation cover, and soil moisture on GNSS-R observable cannot be separated. Empirical and semi-empirical models \cite{CYG_SM_analysis, CYG_SM_ML1, CYG_SM_Semiemp, CYG_SM_SNR, CYG_SM_timeseries,CYG_SM_ANN} relying on external geo-information and SMAP soil moisture estimates have been widely used to infer the impacts from terrain and vegetation. \par
Radar Polarimetry was previously developed to characterize the polarimetric response of a scattered target. This technique has been widely applied to Synthetic Aperture Radar (SAR) systems in both airborne and spaceborne microwave remote sensing missions  to study the polarimetric characteristics of various types of earth surface and vegetation (e.g. \cite{SIR-C, UAV-SAR, AIRSAR}). A similar idea can be applied to GNSS-R: the integration of an additional polarization channel in polarimetric GNSS-R systems has the potential to differentiate surface scattering effects from vegetation in order to retrieve track-wise soil moisture without any ancillary data and provide vegetation characterizations at the same time. The early airborne polarimetric GNSS-R campaigns such as LEiMON \cite{LEimon1,LEimon2}, GRASS \cite{GRASS1}, GLORI \cite{GLORI1,GLORI2,GLORI3,GLORI4}, and the stratospheric balloon experiment BEXUS \cite{BEXUS} have demonstrated the capability of detecting the co-pol scattered GNSS signal and the potential applications of soil moisture, biomass, vegetation water content retrievals and land type classification. The first spaceborne polarimetric GNSS-R was demonstrated by the SMAP-R system \cite{SMAP_R_SV_retrieval} after the Radar failure of SMAP. The first-ever Stokes parameter calibration and retrieval methods \cite{SMAP_R_Spar_retrieval, SMAP_R_calibration} as well as global land remote sensing applications \cite{SMAP_R_science1, SMAP_R_science2, SMAP_R_science3, SMAP_R_science4} have been developed and investigated, while daily spatial-temporal sampling density and observation angle are limited by the special configuration of the SMAP mission. \par
\begin{figure}[!t]
\centering
\includegraphics[width=3.4in]{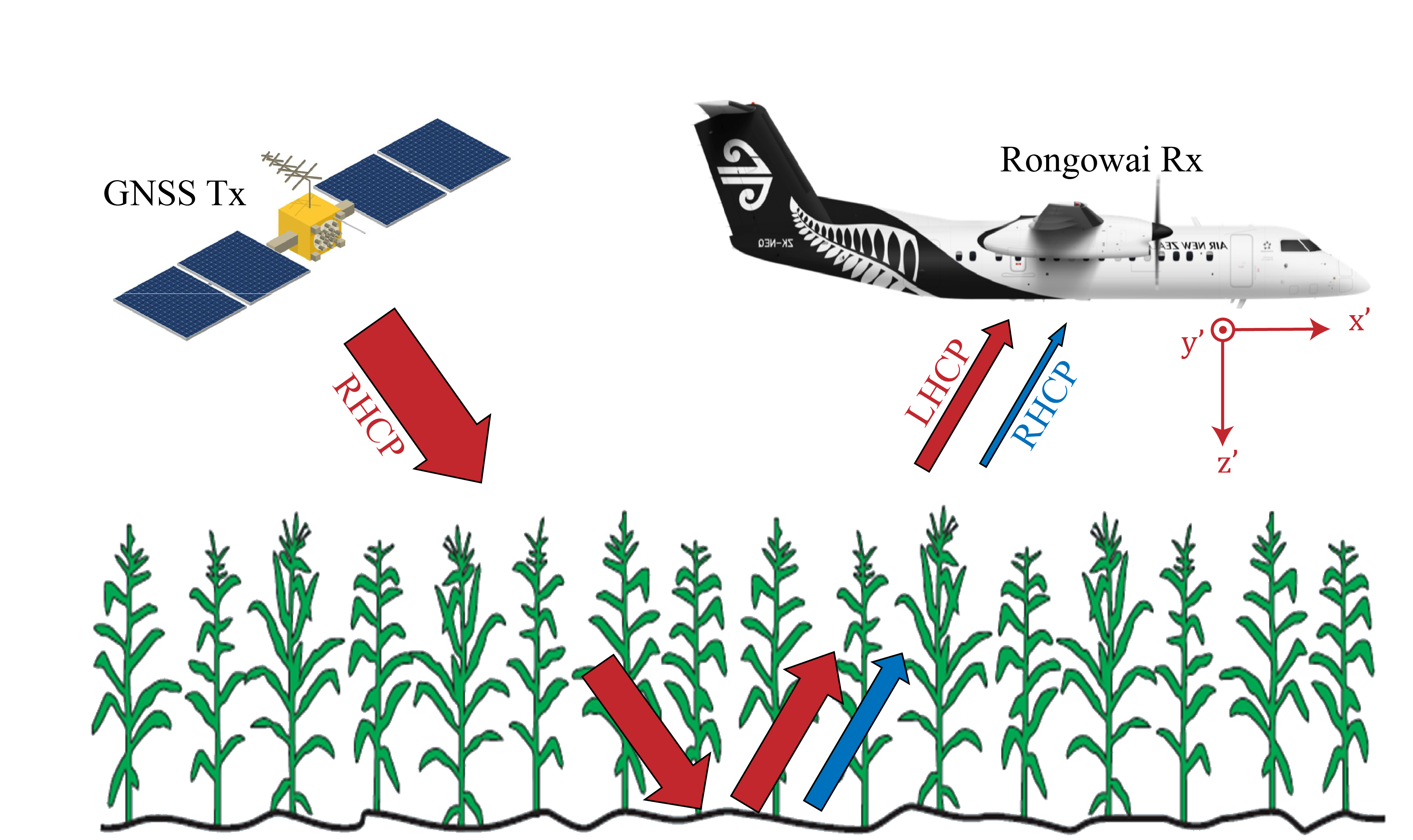}
\caption{Illustration of the idea of land polarimetric remote sensing in the Rongowai mission. The co-pol and cross-pol signals transmitted by GNSS satellites and scattered from rough surface and vegetation are detected by the NGRx system installed on the Air New Zealand Q300 commercial flight}
\label{Intro-graph}
\end{figure}

 
Seeking better coverage, resolution, and geophysical retrieval accuracy to enhance the polarimetric GNSS-R capability, NASA, New Zealand’s Ministry of Business Innovation and Employment, the University of Michigan, the University of Auckland and Air New Zealand initiated an airborne project called “Rongowai” in 2022 \cite{Rongowai}. A newly developed Next Generation GNSS-R Receiver (NGRx) with co- and cross-pol measurement capabilities \cite{NGRx} was installed on a domestic Air New Zealand Q300 airplane in September 2022 to collect data during its daily commercial flights, as shown in Fig. \ref{Intro-graph}.  Rongowai makes near-continuous airborne measurements and generates archival science data products. Compared with traditional GNSS-R systems, the NGRx has the capabilities of processing multi-GNSS-satellite signals simultaneously with a co-pol and cross-pol dual-channel antenna to achieve better spatial coverage and retrieval accuracy. This configuration makes Rongowai the first airborne polarimetric GNSS-R mission that provides daily polarimetric measurements of more than 50,000 samples over different types of land surface with all incidence angles. The open access Level-1 data provides the science community a valuable dataset to investigate polarimetric GNSS-R related techniques and record the long-term climate data in New Zealand \cite{Rongowai_L1_server}. \par

Careful calibration is required to accurately relate the measured data to the properties of Earth’s surface. Calibration converts the raw measurements of a remote sensing system to meaningful physical observables and reduces the errors which can propagate into the higher-level science products. During the development of polarimetric SAR, calibration was typically done by measuring known targets including passive targets like trihedral corner reflector \cite{pol_SAR_passive_cal} and active sources on the ground \cite{pol_SAR_active_cal} which can be modeled accurately. The general idea was to abstract the system potential defects into a set of calibration parameters and solve them with a calibration model and observations from the known targets. A similar approach can be applied to polarimetric GNSS-R calibration. The SMAP-R team first proposed a method to calibrate a polarimetric GNSS-R system with high antenna gain and good cross-polarization isolation \cite{SMAP_R_calibration}, resulting in increased correlation and decreased unbiased root mean square difference (ubRMSD) between a SMAP-R calibrated observable and SMAP/ERA-5 soil moisture. \par

In this paper, we propose a novel calibration method for polarimetric GNSS-R systems with low antenna gain and non-negligible cross-talk between LHCP and RHCP channels by treating inland water bodies and the ocean surface as vicarious calibration targets, and we apply this calibration method to the Rongowai NGRx sensor. Section \hyperref[II]{II} describes a coherent reflection forward model which estimates the coherent power, the power cross-pol ratio model, and the surface dual-polarized effective reflectivity measured by NGRx. In Section \hyperref[III]{III}, the bias in L1 radiometric calibration is determined and corrected by calibrating the L1 power against the inland water coherent reflection model; The antenna cross-pol gain is then estimated and incorporated into calibration algorithm using observations of incoherent scattering from the ocean surface. These calibration steps result in a significant improvement in the consistency between the measurements and the scattering model. Section \hyperref[IV]{IV} presents the surface effective reflectivity retrieval results after calibration, with their statistical distribution and spatial maps illustrated for various surface classifications and with respect to incidence angles. Section \hyperref[V]{V} summarizes the work and suggests the improvements that could be made in future polarimetric GNSS-R calibration.

\section{Theory Background}
\label{II}
 NGRx is calibrated by comparing Rongowai’s observations of specular GNSS reflections from a water surface to a physical model describing the reflections. The water surface is selected as a calibration target due to its homogeneous surface, a significant difference between cross- and co-pol reflectivity, and the distinguishable presence of coherent and incoherent reflections, thus giving us high confidence in the validity of the model.

\subsection{Polarimetric Power Scattering Model for Water body}
\label{IIA}
In the specular direction, the reflected GNSS power received by the GNSS-R receiver includes both coherent and incoherent components \cite{bistatic_radar_eq}:
\begin{equation}
    P = P_{coh} + P_{incoh}
\end{equation}
A matrix form of the Friis transmission formula is used to account for the coherent dual-pol signal reflection from the specular point over water, as given by:
\begin{align}
    \begin{bmatrix}
        P_{L,\mathrm{coh}}\\
        P_{R,\mathrm{coh}}
    \end{bmatrix}
    = \frac{\lambda^2}{(4\pi)^2} & \frac{1}{(R_1+R_2)^2}
    \begin{bmatrix}
        G_{LL} & G_{LR}\\
        G_{RL} & G_{RR}
    \end{bmatrix} \nonumber \\
    &
    \begin{bmatrix}
        \Gamma_{LR} & \Gamma_{LL}\\
        \Gamma_{RR} & \Gamma_{RL}
    \end{bmatrix}
    \begin{bmatrix}
        1\\
        \beta
    \end{bmatrix}
    P_T G_T \cdot \psi
\end{align}

\noindent where $P_{L,\mathrm{coh}}$ and $P_{R,\mathrm{coh}}$ are the received power by the LHCP (cross-pol) and RHCP (co-pol) channels of NGRx; $\lambda$ is the wavelength; $R_1$ and $R_2$ are the signal propagation distances from the specular point to transmitter and receiver respectively; $\Gamma_{pq}$ is the surface reflectivity with q-polarized incident wave and p-polarized scattered field (denoted by L for LHCP, R for RHCP); $P_T G_T$, the product of GNSS transmit power and antenna RHCP gain, is the GNSS RHCP Equivalent Isotropic Radiated Power (EIRP), and $\beta$ is the EIRP cross-pol isolation defined as the fraction of GPS LHCP EIRP with respect to the RHCP EIRP; $\psi$ is the scattering loss due to surface roughness; $G_{pq}$ describes the antenna co-pol ($p$-polarized wave to $p$ channel) and cross-pol ($p$-polarized wave to $q$ channel, which can also be interpreted as channel cross-talk) Gain.  \par

The scattering loss caused by surface roughness is directly related to waterbody surface windspeed \cite{Ulaby_microwave_RS}. The received coherent power at the specular direction is attenuated due to roughness by a factor
\begin{equation}
    \psi = \exp{\left(-4R_a^2 \right)}
\end{equation}
where the Rayleigh Parameter
\begin{equation}
    R_a = 0.5 \pi H_s \frac{\cos{(\theta_i)}}{\lambda}
\end{equation}
depends on specular incidence angle $\theta_i$ and surface significant wave height $H_s$. The CERC model \cite{loria_model} relates $H_s$ to surface wind speed by:
\begin{align}
    H_s = \frac{U_A^2}{g} & 0.283\tanh{\left[ 0.53 \left( \frac{g\cdot d}{U_A^2} \right)^{ \frac{3}{4} } \right]} 
    \nonumber \\
    & \cdot \tanh{\left\{  \frac{0.00565 \left( \frac{g\cdot F}{U_A^2}   \right) ^{\frac{1}{2}} } 
    {\tanh{\left[ 0.53 \left(\frac{g\cdot d}{U_A^2} \right)^{\frac{3}{4}} \right]}}  \right\}}
\end{align}
\noindent In the above equation, $U_A=0.7 U_{10}^{(1.23)}$, where $U_{10}$ is the windspeed 10 meters above the water surface. $g$ is the gravitational acceleration ($\mathrm{m}/\mathrm{s}^2$), $d$ is the water depth (m), and $F$ is the Fetch length (m). \par

\begin{figure}[!t]
\centering
\includegraphics[width=3.4in]{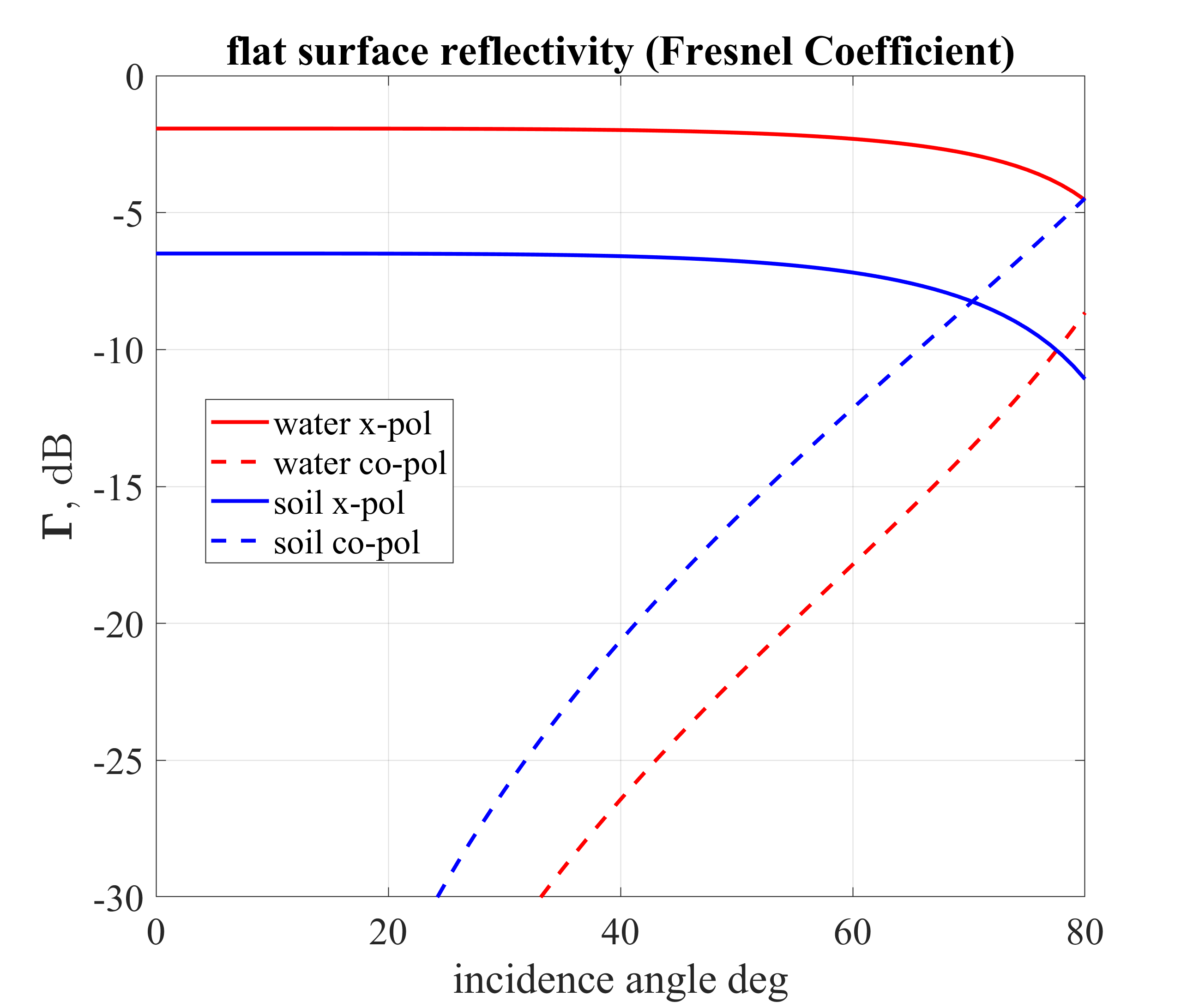}
\caption{Fresnel coefficient of $ \Gamma_{\text{co-pol}}$ and $ \Gamma_{\text{x-pol}}$ of fresh water (red) with $\epsilon=80.97-8.44j$ and loam type of soil (blue) with 10\% volumetric moisture content with $\epsilon=7.72-1.04j$, plotted versus incidence angle}
\label{soil-water-Fresnel}
\end{figure}

The surface Fresnel reflectivities $\Gamma_{pq}$ in circular polarization LR/RL (cross-pol, denoted as x-pol) and LL/RR (co-pol) are:
\begin{equation}
    \Gamma_{\text{x-pol}} = \Gamma_{LR} = \Gamma_{RL} = \left | \frac{1}{2}(\mathfrak{R}_{VV}-\mathfrak{R}_{HH}) \right |^2
\end{equation}
\begin{equation}
    \Gamma_{\text{co-pol}} = \Gamma_{LL} = \Gamma_{RR} = \left | \frac{1}{2}(\mathfrak{R}_{VV}+\mathfrak{R}_{HH}) \right |^2
\end{equation}
The Cross- and Co-pol naming is based on the matchup of the transmit and receiver antenna polarizations. here:
\begin{equation}
    \mathfrak{R}_{VV} = \frac{\epsilon \cos{\theta_i} - \sqrt{\epsilon-\sin^2{\theta_i} } } {\epsilon \cos{\theta_i} + \sqrt{\epsilon-\sin^2{\theta_i} } }
\end{equation}
\begin{equation}
    \mathfrak{R}_{HH} = \frac{ \cos{\theta_i} - \sqrt{\epsilon-\sin^2{\theta_i} } } {\cos{\theta_i} + \sqrt{\epsilon-\sin^2{\theta_i} }}
\end{equation}
is a function of incidence angle $\theta_i$ and surface relative complex permittivity $\epsilon$. For a water surface, $\epsilon$ can be estimated by a microwave ocean dielectric model \cite{Water_eps_model} as a function of salinity, temperature, and frequency. In Fig. \ref{soil-water-Fresnel}, the cross-pol and co-pol Fresnel Coefficients ($\Gamma_{\text{x-pol}}$ and $\Gamma_{\text{co-pol}}$ ) of fresh water  ($\epsilon=80.97-8.44j$) and loam soil with 10\% volumetric moisture content ($\epsilon=7.72-1.04j$) \cite{Ulaby_microwave_RS}, are plotted versus incidence angle. Note that the co-pol reflectivity of water is much smaller than that of moist soil.   \par

Similar to the coherent case, the incoherent power received in the specular direction over a water surface can be described by the bistatic Radar equation \cite{zavorotny_KAGO}, also in matrix form as:
\begin{align}
    \begin{bmatrix}
        P_{L,\mathrm{incoh}}\\
        P_{R,\mathrm{incoh}}
    \end{bmatrix}
    = \frac{\lambda^2}{(4\pi)^3} & \frac{1}{R_1^2R_2^2}
    \begin{bmatrix}
        G_{LL} & G_{LR}\\
        G_{RL} & G_{RR}
    \end{bmatrix} \nonumber \\
    &
    \begin{bmatrix}
        \sigma_{LR} & \sigma_{LL}\\
        \sigma_{RR} & \sigma_{RL}
    \end{bmatrix}
    \begin{bmatrix}
        1\\
        \beta
    \end{bmatrix}
    P_T G_T 
\end{align}
\noindent All variables in the equation above vary with the observation spatial geometry, however, for simplification and with negligible loss in accuracy, they can be estimated and applied as constants in the specular direction \cite{cygnss_handbook, SP_coherency}. Compared with (2), the effective reflectivity matrix $\mathbf{\Gamma} \cdot \psi$ is replaced by the BRCS $\mathbf{\sigma}$ matrix. Note that in the specular direction, $\sigma_{pq} \approx \sigma_{pq}^0\cdot A $, where $\sigma_{pq}^0$ is the NBRCS and $A$ is the scattering area. \par

\subsection{Power Cross-pol Ratio}
\label{IIB}
Assuming no GNSS cross-pol EIRP, i.e. the GNSS satellites transmit pure RHCP signals and $\beta=0$, the specular power cross-pol ratio for coherent reflections is:
\begin{equation}
    \frac{P_{R,\mathrm{coh}}}{P_{L,\mathrm{coh}}} = \frac{ G_{RL} \Gamma_{LR} + G_{RR} \Gamma_{RR} } 
    {G_{LL}\Gamma_{LR} + G_{LR} \Gamma_{RR}}
\end{equation}
A receive antenna with decent cross-pol isolation satisfies $G_{LL} \gg G_{LR}$ and $G_{RR} \gg G_{RL}$ . For reflections from water surfaces,  $\Gamma_{LR} \gg \Gamma_{RR}$, so the cross-pol leakage term $G_{LR}\Gamma_{RR}$ in the LHCP channel can be ignored. However in the RHCP channel, the antenna cross-pol coupling $ G_{RL} \Gamma_{LR}$ and co-pol gain $G_{RR} \Gamma_{RR}$ can be comparable in magnitude. Then (11) can be simplified to
\begin{equation}
    \frac{P_{R,\mathrm{coh}}}{P_{L,\mathrm{coh}}} = \frac{G_{RL}}{G_{LL}} + \eta \cdot \frac{\Gamma_{RR}}{\Gamma_{LR}}
\end{equation}
where $\eta = \frac{G_{RR}}{G_{LL}} \approx 1$ for NGRx dual-pol receive antenna. Equation (12) implies that the power cross-pol ratio measured by the receiver is the sum of the antenna cross-pol ratio and the surface reflectivity cross-pol ratio, highlighting the fact that the antenna cross-pol ratio must be calibrated carefully in order to retrieve surface co-pol reflectivity accurately. \par

Similar to (12), the incoherent power cross-pol ratio in the specular direction is:
\begin{equation}
    \frac{P_{R,\mathrm{incoh}}}{P_{L,\mathrm{incoh}}} = \frac{G_{RL}}{G_{LL}} + \eta \cdot \frac{\sigma_{RR}^0}{\sigma_{LR}^0}
\end{equation}
 Based on the KA-GO model \cite[pp. 88]{tsang1985theory}, \cite{Johnson_KAGO},  $\frac{\sigma_{RR}^0}{\sigma_{LR}^0} = \frac{\Gamma_{RR}}{\Gamma_{LR}}$ in the specular direction for a homogeneous surface (e.g. ocean), since the scattering cross section for all polarizations is proportional to the same function of surface roughness.

\begin{figure}[!t]
\includegraphics[width=3.4in]{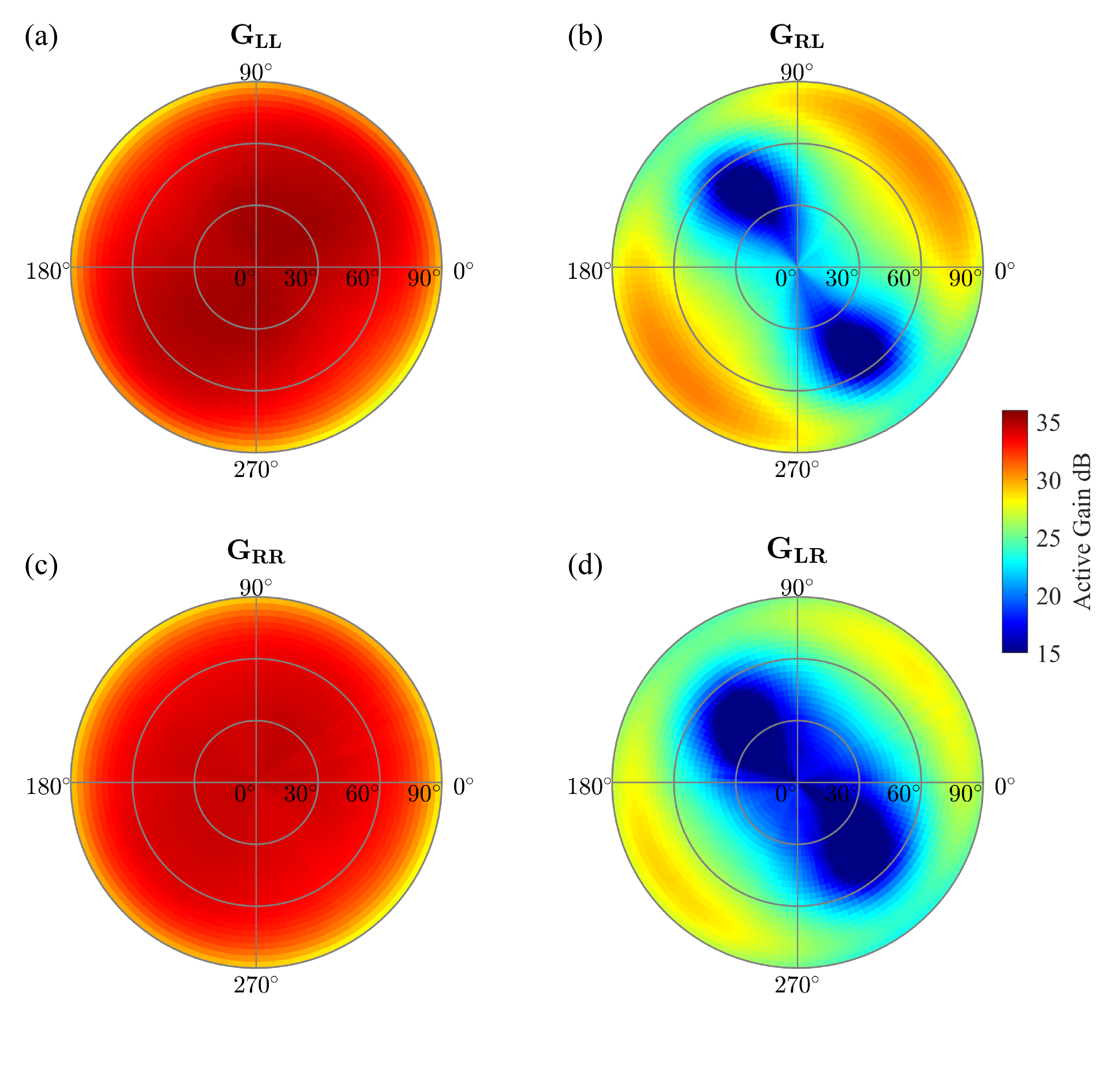}
\caption{Full pattern of the active dual-pol antenna in dB scale at 1575 MHz (a) $G_{\mathrm{LL}}$ (b) $G_{\mathrm{RL}}$ (c) $G_{\mathrm{RR}}$ (d) $G_{\mathrm{LR}}$ measured in the anechoic chamber from ElectroScience Lab, The Ohio State University. The  antenna polar coordinate defines radius as the off-boresight angle and polar angle as the azimuth angle.} 
\label{antenna measured pattern}
\end{figure}

\subsection{Surface Effective Reflectivity Retrieval}
\label{IIC}
 Assuming 100\% coherence, the surface effective reflectivity can be retrieved by invoking symmetries in the polarization dependence of the surface reflectivity and inverting the matrix in (2):

\begin{align}
    \begin{bmatrix}
        \hat{\Gamma}_{LR} \\
        \hat{\Gamma}_{RR}
    \end{bmatrix}
    =
    \frac{(4\pi)^2}{\lambda^2} & \frac{ (R_1+R_2)^2}{ P_TG_T}  
    \begin{bmatrix}
        1 & \beta \\
        \beta & 1
    \end{bmatrix} ^{-1} \nonumber \\
    &
    \begin{bmatrix}
        G_{LL} & G_{LR}\\
        G_{RL} & G_{RR}        
    \end{bmatrix} ^{-1}
    \begin{bmatrix}
        P_L \\
        P_R
    \end{bmatrix}
\end{align}
where $P_L$ and $P_R$ are the power measured by the LHCP and RHCP channels respectively. The retrieved effective reflectivity is the product of the Fresnel coefficient, the scattering loss due to roughness, and the extinction loss due to vegetation, or:
\begin{equation}
    \hat{\Gamma}_{pq} = \Gamma_{pq}\psi \gamma
\end{equation}
with
\begin{equation}
    \gamma = \exp{(-2\cdot \tau_v \sec{\theta_i})}
\end{equation}
describing the attenuation due to vegetation in terms of Vegetation Optical Depth (VOD) $\tau_v$ and and incidence angle $\theta_i$. Note that $\gamma$ equals 1 on a vegetation-free, open water surface and so is ignored in (2).

\begin{figure}[!t]
\centering
\includegraphics[width=3.3in]{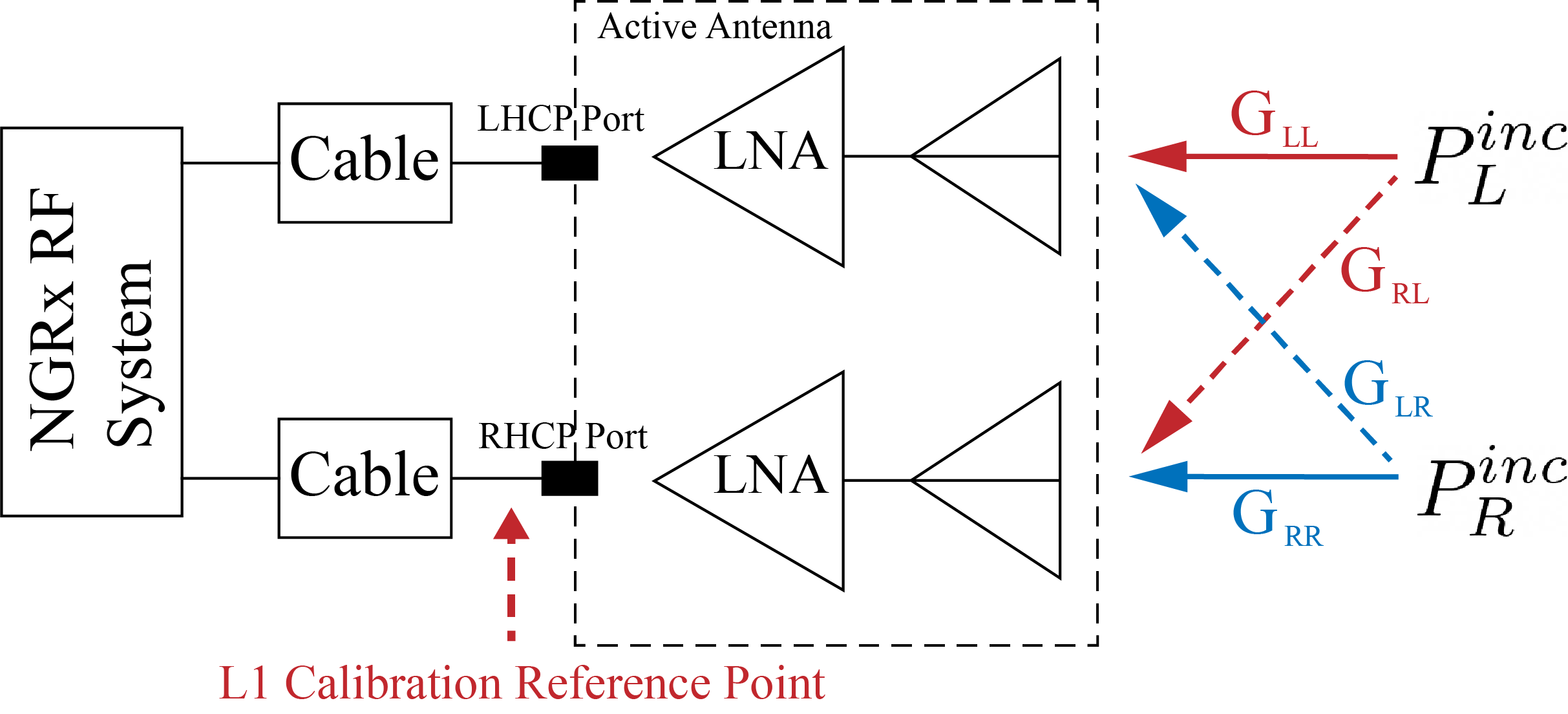}
\caption{System diagram of the Rongowai front end. The dual-pol antenna has LHCP and RHCP ports connected to the NGRx system with two long co-axial cables. The L1 Calibration reference point is labeled. The incident signal's power flow due to the antenna co-pol gain and cross-pol gain are illustrated.}
\label{system-diagram}
\end{figure}


\section{Calibration Methodology}
\label{III}
Retrieval of surface effective reflectivity requires accurate evaluation of each term in (14). Pre-launch tests provide prior information about L1 power conversion and the full antenna pattern, however they might be biased or corrupted after on board installation. As a result, terms in the end-to-end L1 calibration expression are revisited using airborne measurements. L1 calibration is done by comparing the on board behavior of the polarimetric measurement over water surfaces (ocean and inland lake) with the physical model of the same scenario described in Section II A and B. We use data obtained between 4/16/2023 and 11/13/2023 with 2 ms coherent integration time to demonstrate the calibration method and derive corresponding results.

\subsection{Pre-launch Calibration}
\label{IIIA}
Two dual-pol stacked patch GPS active antennas (P/N: 42G1215RL-AA-5SSF-1, produced by Antcom Corporation \url{https://www.antcom.com/}) are installed on the Rongowai Q300 aircraft. The zenith antenna receives GPS navigation information, while the nadir antenna is used for science measurements. The full patterns of the antenna's active gain at all elevation and azimuth angles were measured in an anechoic chamber in the ElectroScience Lab, Ohio State University. The measured patterns are shown in Fig. \ref{antenna measured pattern} and plotted in the antenna 2-D polar coordinate which is defined by the off-boresight angle as the radius, and the azimuth as the polar angle. The pattern measurement has $3^{\circ}$ resolution in both off-boresight and azimuth dimensions. The nadir antenna was installed on the Q300 airplane with 0 azimuth pointing towards the nominal flight heading direction $x^\prime$, 90$^{\circ}$ towards $y^{\prime}$, with the Q300 airplane local coordinate system $x^{\prime}y^{\prime}z^{\prime}$ defined in Fig. \ref{Intro-graph}.   \par

Fig. \ref{system-diagram} illustrates the system front end functional blocks. The incident GNSS signal described by LHCP and RHCP polarizations is received by the active antenna and output at its LHCP and RHCP ports, with both co-pol and cross-pol gain considered. The output signals are then propagated through two well-isolated but lossy Coaxial cables and input into the NGRx flight unit for demodulation, sampling, cross-correlation, and generation of Level 0 raw-counts Delay Doppler Maps (DDMs). The L1 power DDM represents the power at the reference point labeled in Fig. \ref{system-diagram} in units of watts. The L1 power conversion function is developed to map the raw counts to power and was determined prior to installation on the Q300 airplane \cite{Rongowai_ATBD}.

\begin{figure}[!t]
\centering
\includegraphics[width=3.1in]{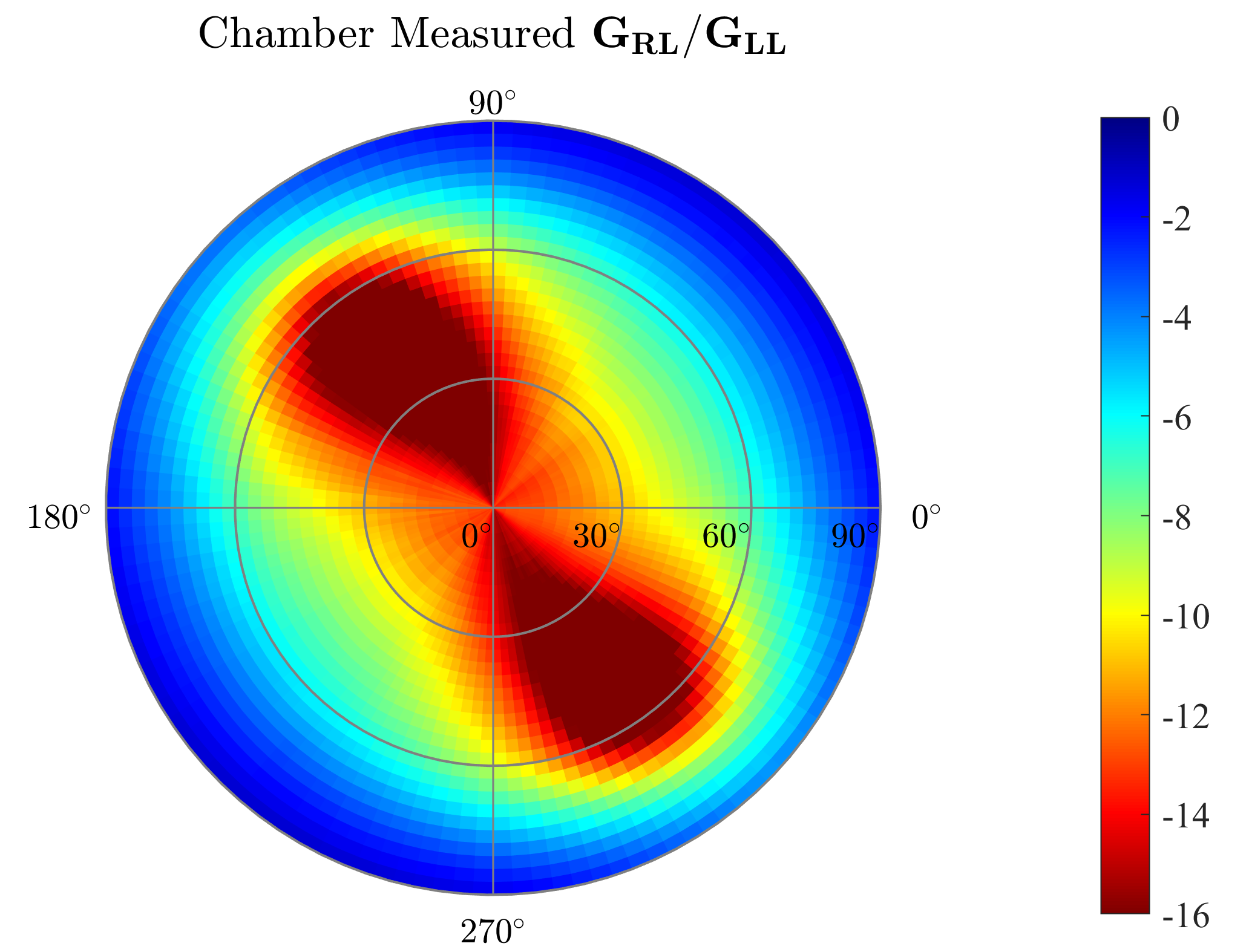}
\caption{Chamber measured antenna cross-pol ratio pattern $G_{RL}/G_{LL} $ (dB) before launch, plotted in the antenna coordinate.}
\label{xpol before cal}
\end{figure}

\begin{figure}[!t]
\centering
\includegraphics[width=3.3in]{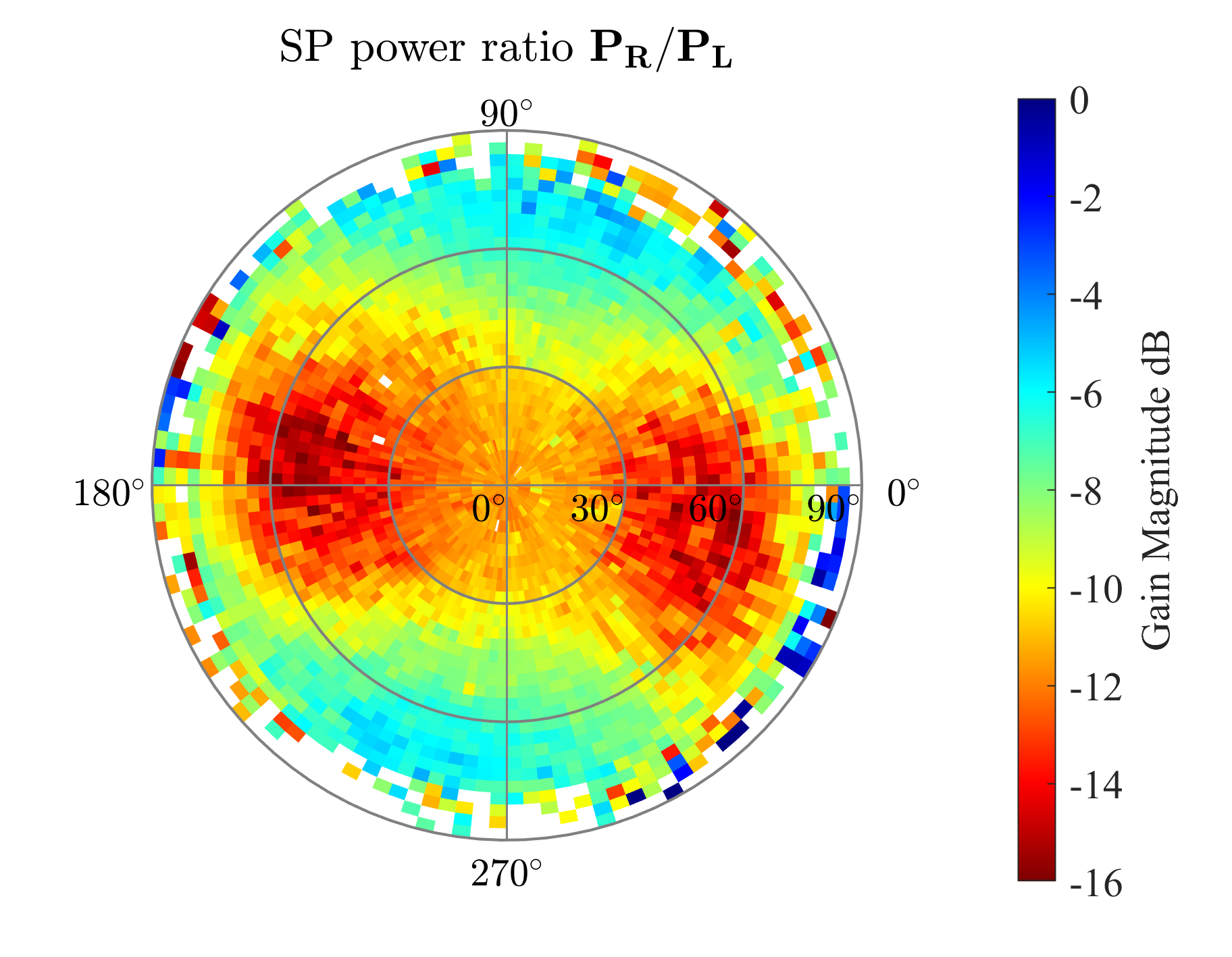}
\caption{Binned and averaged measured specular power cross-pol ratio $P_R/P_L$ (dB) obtained from incoherent ocean surface with $3^\circ$ resolution in azimuth and elevation, plotted in the antenna coordinate. }
\label{ocean incoherent power xpol ratio}
\end{figure}

\subsection{Antenna Pattern Rotation}
\label{IIIB}
Below 65$^{\circ}$ incidence angle, the RHCP/LHCP Fresnel Coefficients ratio is under -15 dB. According to (12) and (13), specular $\frac{P_{R}}{P_{L}} \approx \frac{G_{RL}}{G_{LL}}$ for viewing angles at which the antenna cross-pol ratio is much higher than the Fresnel Coefficients cross-pol ratio, so the measured specular power cross-pol ratio should display the antenna cross-pol ratio pattern $G_{RL}/G_{LL}$ in Fig. \ref{xpol before cal}. \par

\begin{figure}[!t]
\centering
\includegraphics[width=3.3in]{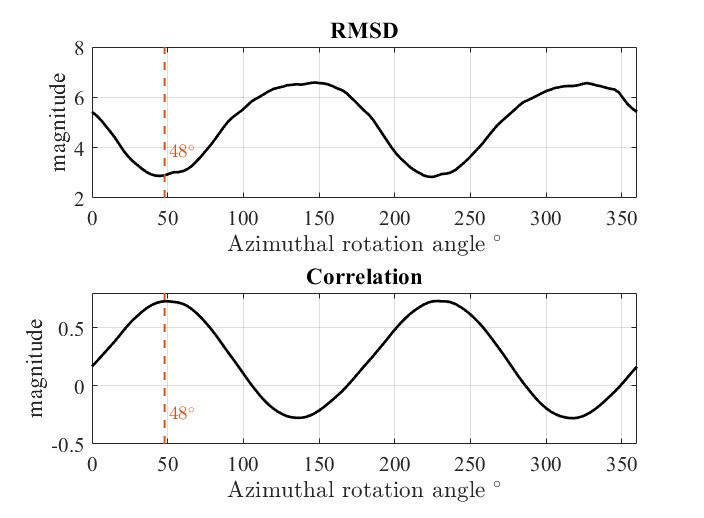}
\caption{RMSD and Pearson correlation between pre-launch chamber measured antenna cross-pol ratio pattern and measured power cross-pol ratio from incoherent ocean surface. Minimum RMSD and Maximum correlation are achieved simultaneously at $48^\circ$ azimuth rotation (counter-clockwise).}
\label{antenna rotation rmsd and corr}
\end{figure}

The measured power cross-pol ratio $P_R/P_L$ obtained from ocean surface (1,423,997 samples), binned and averaged within 3$^{\circ}$ off-boresight and azimuth angle window, is plotted in the antenna polar coordinate in Fig. \ref{ocean incoherent power xpol ratio}. Samples with LHCP channel SNR $<$ 3 dB are excluded. A clear azimuth rotation bias is observed. The exact cause or causes of the rotation is unknown, but may be the result of a mechanical rotation during aircraft installation, or it may be a modification of the antenna pattern due to the aircraft metal body behind the antenna acting as an azimuthally asymmetric ground plane. To compensate for the rotation, the pre-launch measured antenna pattern is rotated in azimuth while its RMSD and correlation with respect to the measured power cross-pol ratio are computed , as shown in Fig. \ref{antenna rotation rmsd and corr}. The $180^{\circ}$ periodicity in both RMSD and correlation curves is caused by the symmetric antenna cross-pol pattern. The azimuth bias angle is determined to be 48$^{\circ}$ by finding that RMSD and correlation are minimized and maximized simultaneously. This azimuth bias correction is applied to antenna co-pol and cross-pol gain in all the following sections of this paper.

\subsection{Power Measurement Bias Calibration}
\label{IIIC}
\begin{figure}[!t]
\centering
\includegraphics[width=3.45in]{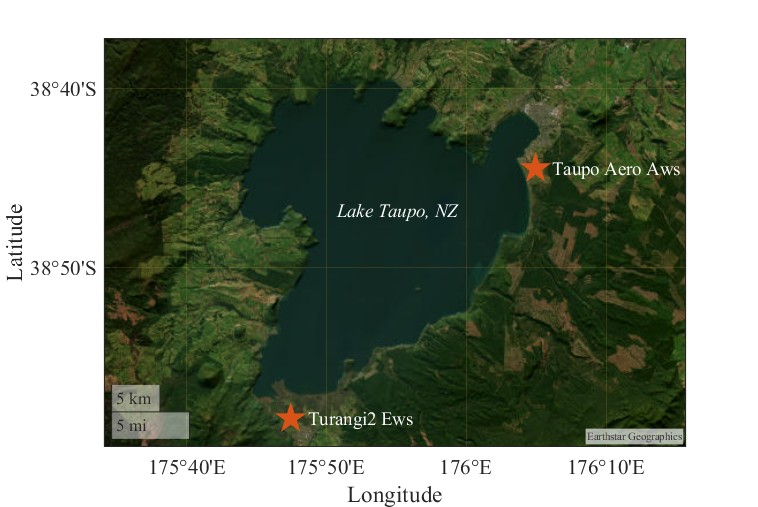}
\caption{Taupo satellite map (source: Google Earth). Wind speed information are obtained from the two weather stations near the lake shore labeled as red star.}
\label{Taupo map}
\end{figure}

Power measurements obtained from Lake Taupo, the largest inland lake in New Zealand, are compared with the model described in section \hyperref[IIA]{IIA} under the coherent scenario to calibrate the L1 power. Taupo's satellite map is shown in Fig. \ref{Taupo map}, with two weather stations Taupo Aero AWS and Turangi2 EWS labeled, where the lake surface windspeed was recorded hourly during the calibration period. The windspeed data recorded at Turangi2 EWS and Taupo Aero AWS are provided by NIWA Cliffo Service \url{https://cliflo.niwa.co.nz/} and MetService \url{https://www.metservice.com/} respectively. A coherence detector based on the L1 power DDM \cite{coherence_detector}, which determines the coherence state by comparing the DDM delay waveform with GNSS signal ambiguity function, is applied to select data with strong coherence state. Data with the specular point less than 300 meters away from the Taupo coastline are discarded to prevent land contamination. The following filtering criteria are also applied: data is selected if the off-boresight angle is less than $65^{\circ}$ since data with high angles is subject to multi-path; the LHCP channel SNR is larger than 4 dB; the angle between two consecutive measurement's velocity vector is smaller than $0.01^{\circ}$ to exclude data with large and varying roll-pitch-yaw due to the flight motion. In the end, 2659 samples of high quality obtained from Lake Taupo are available for calibration. \par

\begin{table}[!t]
\centering
\caption{Taupo Geographic Parameters Estimated for Model}
\label{Taupo parameter} 
\begin{tabular}{|c||c||c|} 
\hline
\textbf{Parameter} & \textbf{Value} & \textbf{Comment} \\
\hline
$U_{10}$ & 1.71 m/s & Mean windspeed at 10m height \\
D & 91 m & Mean water depth of Taupo \\
F & 5 km & Mean fetch \\
T & $10^{\circ}C$ & Winter mean water temperature \\
S & 0 & Fresh water Salinity \\
\hline
\end{tabular}
\end{table}

The coherent model described in (2)-(9) is applied to estimate the received specular power at the calibration reference point. The observation geometry information is computed by the specular searching algorithm \cite{Rongowai_SP} and is available at the open access L1 data archive. Antenna gain per sample is obtained by interpolating the pre-launch pattern to the actual off-boresight and azimuth angles. GPS EIRP information is obtained from the static EIRP Lookup Table derived during the CYGNSS mission \cite{GPS_static_EIRP}. GPS cross-pol EIRP is assumed to be zero. The coherent reflection model also requires geographic information about Lake Taupo in order to estimate the scattering loss due to surface roughness and water dielectric constant. It is not possible to measure these parameters accurately per sample, so the averaged values of them over the calibration period, as summarized in table \ref{Taupo parameter}, are used instead to estimate the mean dielectric constant and scattering loss. The Fetch length of each sample is calculated as the distance between sample and coastline in the upwind direction, where the wind direction of each sample is obtained by interpolating the weather stations data. The mean windspeed is obtained by interpolating the weather stations data with each sample time and averaging them. Samples with strong coherence state determined by the coherence detector \cite{coherence_detector} have a mean windspeed of 1.7 m/s, compared with 3 m/s mean windspeed from all samples with coherent, incoherent, and mixed coherent states. The reduction in windspeed suggests a stronger coherence state of the selected samples. \par

\begin{figure}[!t]
\centering
\includegraphics[width=3.3in]{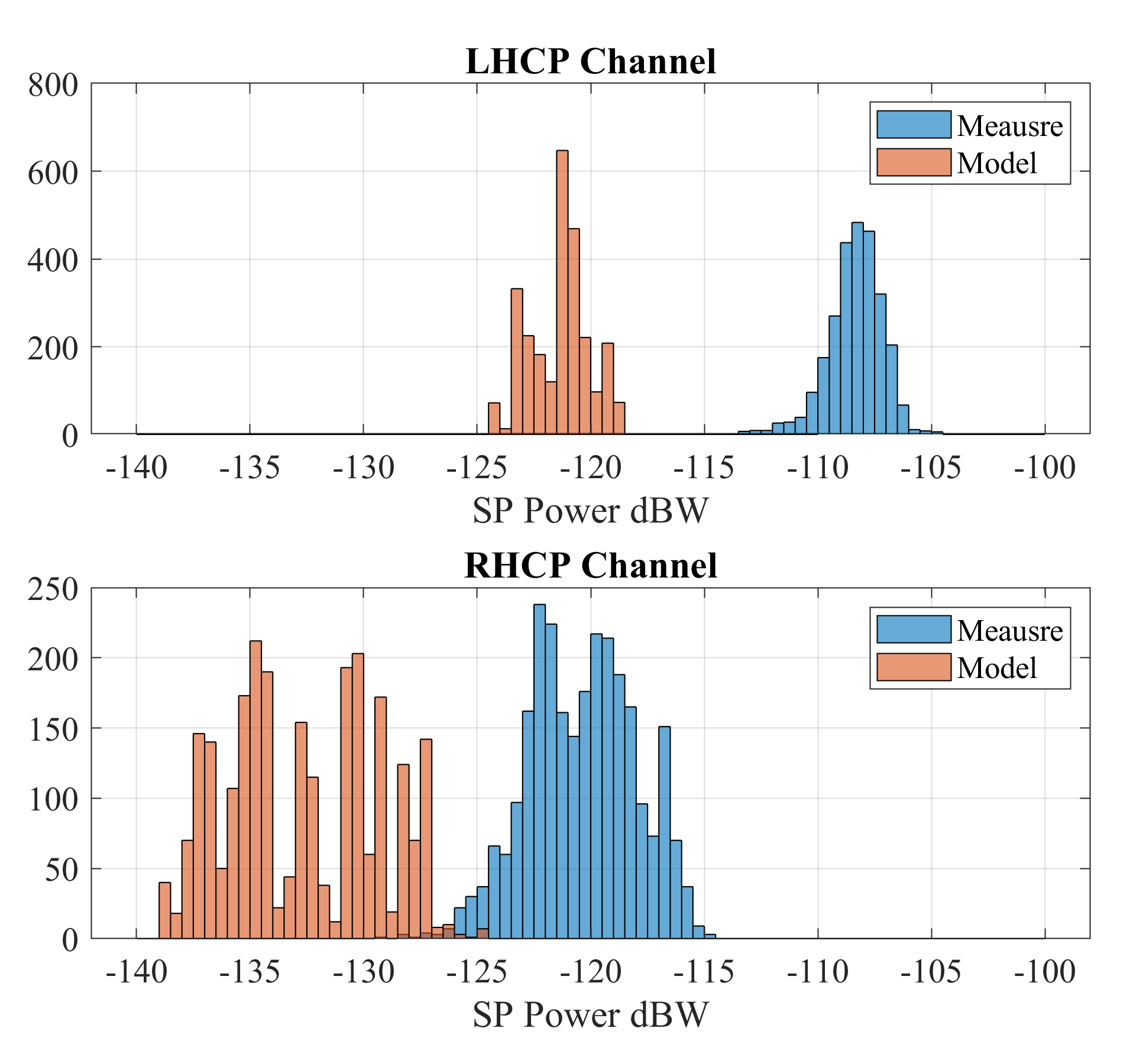}
\caption{Measured DDM specular power (blue) vs modeled power (orange) in dBW. Upper: LHCP Channel; Lower: RHCP Channel}
\label{SP power histogram}
\end{figure}

\begin{figure}[!t]
\centering
\includegraphics[width=3.3in]{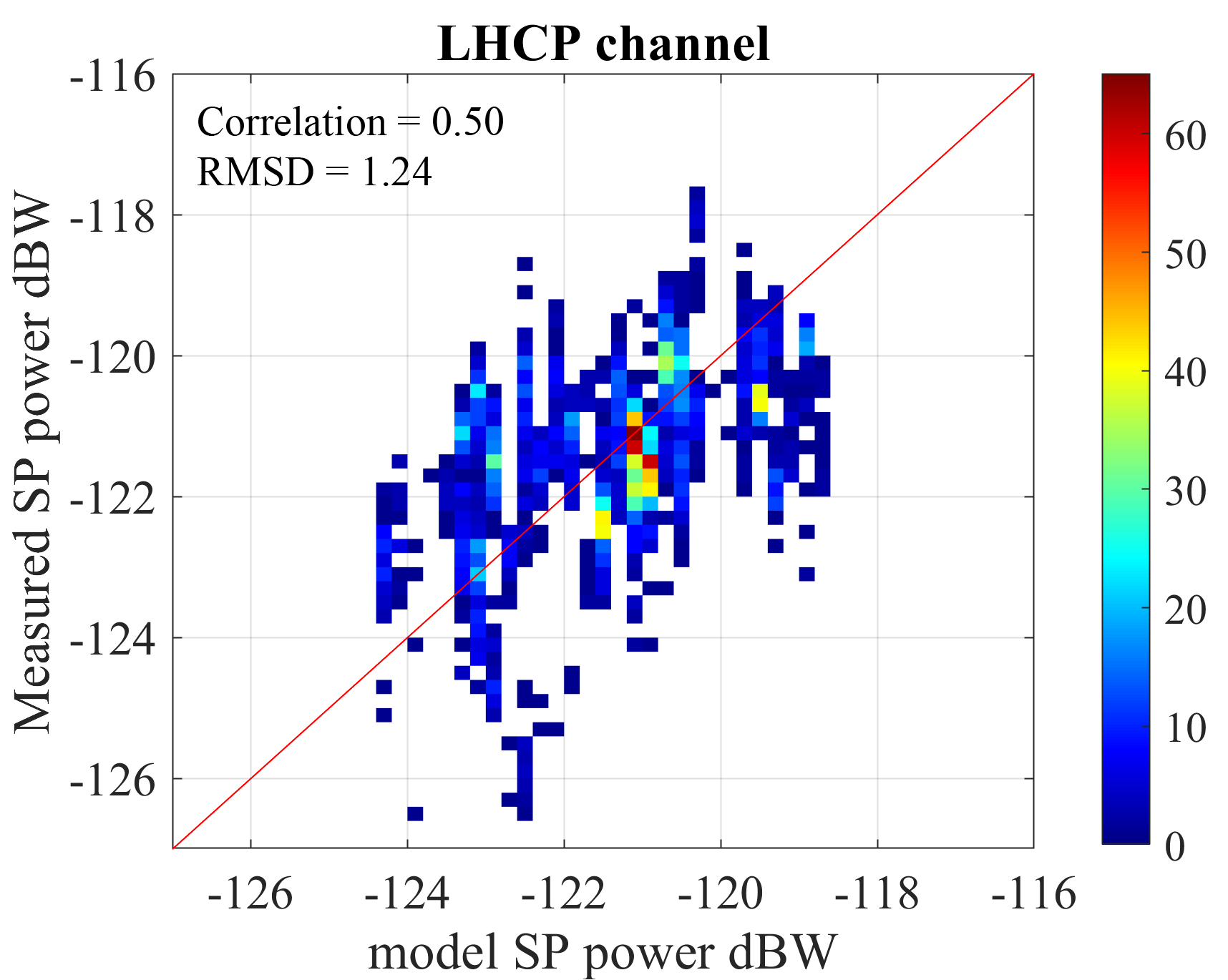}
\caption{scatter density plot of measured SP power after PCF correction vs modeled SP power in dBW. Red line indicates the 1:1 fit.}
\label{power debias model vs measure scatter plot}
\end{figure}

Histograms of the measured LHCP and RHCP specular power are shown in Fig. \ref{SP power histogram} in blue, compared with estimated power by the coherent model of the same scenario in orange. The measured power is approximately 10 dB higher than the model prediction. If left uncorrected, this systematic scale bias in power measurement will lead to overestimated surface reflectivity or NBRCS. The scale bias in the power measurement possibly results from a bias in pre-launch raw counts to L1 power conversion function calibration. \par

This bias motivates the inclusion of a distortion matrix to (2) and (10) to characterize the non-ideal effects in the receiver, which is similar to the receiver distortion matrix used in polarimetric SAR calibration \cite{pol_SAR_passive_cal}. In general,
\begin{equation}
    \begin{bmatrix}
        \Tilde{P_L} \\
        \Tilde{P_R}
    \end{bmatrix}
    = K
    \begin{bmatrix}
        1 & \delta_{LR} \\
        \delta_{RL} & \kappa_{RR}
    \end{bmatrix}
    \begin{bmatrix}
        P_L \\
        P_R
    \end{bmatrix}
\end{equation}
where $P_L$ and $P_R$ are the power at the L1 calibration reference point, while $\Tilde{P_L}$ and $\Tilde{P_R}$ are measured  power at the NGRx RF system; $K$ and $\kappa_{RR}$ are the radiometric power correction factor and the channel imbalance, while $\delta_{LR}$ and $\delta_{RL}$ are channel cross-talks. The NGRx system is designed to have balanced gains on dual-pol channels and no cross-pol coupling, so the distortion matrix become an identity matrix.The power correction factor $K$ is determined by the scale $Q$ that minimizes the RMSD between the measured and modeled power scaled by $Q$ dB with expressions in dBW:

\begin{align}
    K &=  \operatorname*{argmin}_{Q} 
    \left[ \sqrt{\frac{1}{N} \sum_i 
    \left( \hat{P}_{i,\mathrm{dBW}} -  P_{i, \mathrm{dBW}} - Q \right)^2 }   \right] \nonumber \\
    &= \frac{1}{N} \sum_i \left( \hat{P}_{i,\mathrm{dBW}} -  P_{i, \mathrm{dBW}} \right)
\end{align}

where $P_{i, \mathrm{dBW}}$ is the $i^{\mathrm{th}}$ measured power sample, while $\hat{P}_{i,\mathrm{dBW}}$ is the corresponding modeled power. Finding the minimum of RMSD is mathematically equivalent to finding the mean bias between measured and modeled power. The power correction factor $K$ is calculated to be -13.03 dB for the LHCP channel, and a minimum RMSD of 1.24 between model and measurement is achieved. The scatter density plot is shown in Fig. \ref{power debias model vs measure scatter plot}, where the red solid line indicates the 1:1 fit of the modeled and measured results. The model and debiased measurement agrees well with each other, with a Pearson correlation coefficient of 0.50.  \par

Note that only the LHCP channel is used to calibrate $K$, since the power measured in the RHCP  
channel strongly depends on the antenna cross-pol gain, especially on water surfaces where $\Gamma_{RR}/\Gamma_{LR}$ is small, as shown in (12). However, the on board cross-pol gain might be corrupted and changed from its pre-launch pattern (for example, the azimuth rotation described in section \hyperref[IIIB]{IIIB}) due to its interaction with the aircraft metal body, so the RHCP channel power cannot be calibrated correctly without first calibrating the antenna cross-pol pattern.

\begin{figure*}[t!]
\centering
\includegraphics[width=6.5in]{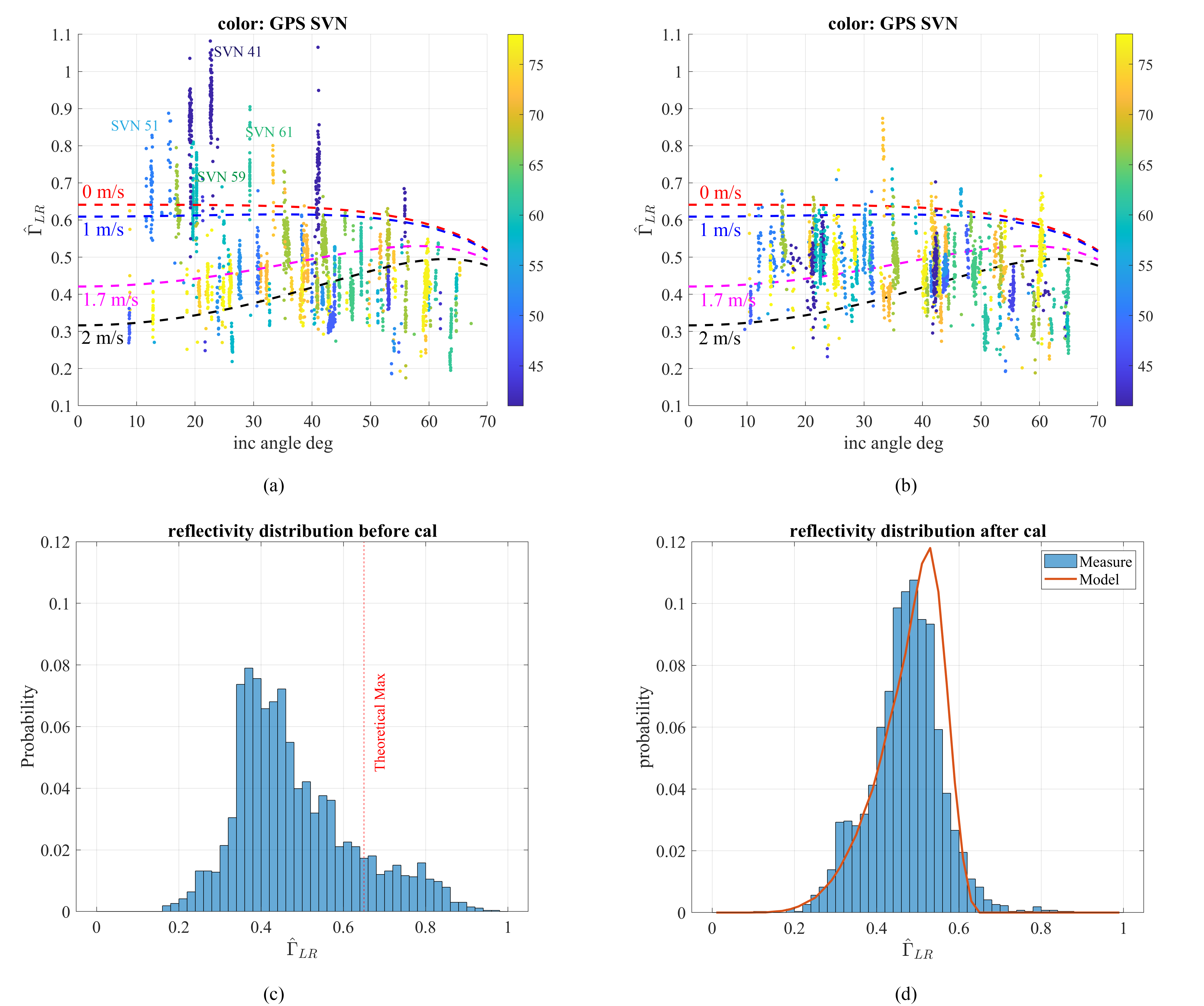}
\caption{Reflectivity calibration over Lake Taupo surface: (a) scatter plot of retrieved cross-pol (LR) effective reflectivity vs incidence angle before calibration. Color indicates the SVN. Dash lines show modeled effective reflectivity with different windspeed labeled beside; (b) same as (a) after calibration; (c) distribution of retrieved cross-pol (LR) reflectivity before calibration; (d) same as (c) after calibration. Red curve suggests the simulated result.}
\label{refl cal}
\end{figure*}

\subsection{Water Reflectivity Calibration}
\label{IIID}
The surface effective reflectivity of Lake Taupo can be determined by (14) after calibrating the power measurement bias, and its distribution is shown in Fig. \ref{refl cal} (c). The theoretical max, representing the nadir direction reflectivity with 0 m/s surface windspeed, is indicated by the dashed red line in the same plot. However, a large number of samples are higher than it. This problem is examined by plotting the retrieved surface effective reflectivity versus incidence angle from 0 to $65^{\circ}$ in Fig. \ref{refl cal} (a). The modeled surface effective reflectivity with 0, 1 m/s, 1.7 m/s (average windspeed), and 2 m/s are estimated from (3)-(9) and plotted versus incidence angles in dashed curves. Each data point's color indicates its GPS Space Vehicle Number (SVN). Each group of data that appears as a vertical bar is obtained from a consecutive time series of measurements through which the observing geometries are about the same, and the variation in reflectivity is mainly caused by changes in the wind-induced surface roughness. In this plot, it is clear that a few groups of effective reflectivity measured from Block IIR GPS satellites (SVN41, SVN51, SVN59, SVN61) are overestimated, showing a much higher reflectivity than the major population of the data. The problem is very likely caused by the GPS static EIRP dataset that was created in 2018 \cite{GPS_static_EIRP}: the GPS EIRP is the product of the GPS transmit power and its antenna gain with a fixed pattern in azimuth and elevation angles, so the GPS transmit power might be changed and deviated from the original dataset. Furthermore, the bias in transmit power should be consistent at different incidence angles. Here, attempts are made to calibrate the GPS transmit power empirically with the data that is filtered for calibration in section II C. Firstly, two assumptions are made:
\begin{itemize}
    \item On the lake surface, the RHCP-LHCP channel leakage in (2) is much smaller than the LHCP-LHCP co-pol signal in the LHCP channel, so the leakage term can be ignored. Then, the LHCP channel measurement and its model can be treated as a traditional GNSS-R with single polarization channel with retrieved surface effective reflectivity being inversely proportional to the GPS EIRP;
    \item Data measured from each GPS space vehicle within 10-20, 20-30, 30-40, 40-50, 50-60 degrees incidence angle consists of around 50 samples, and the surface condition of each measurement has uncertainty in its windspeed. Assume that one of the measurements with the max reflectivity corresponds to the lowest windspeed near 0.
\end{itemize}

 Then, the bias in GPS transmit power per SVN within each 10 degrees window of incidence angle can be estimated by:
\begin{equation}
    \Delta P_{T, \mathrm{dB}} = \Gamma_{LR, \mathrm{dB}}(\theta_i) - 
    \max \left\{ \hat{\Gamma}_{LR, \mathrm{dB} } \right\} 
\end{equation}
where $\Gamma_{LR, \mathrm{dB}}(\theta_i)$ is the theoretical max cross-pol reflectivity at incidence angle $\theta_i$, calculated from (6), and $\hat{\Gamma}_{LR, \mathrm{dB}}$ is the measured reflectivity. The final results of $\Delta P_{T,\mathrm{dB}}$ are summarized in table \ref{EIRP_adj} in appendix A. Finally, the transmit power bias of each SVN is obtained by averaging $\Delta P_{T, \mathrm{dB}}$ from each incidence angle, weighted by the number of samples. A negative value means the SVN's transmit power is underestimated.

\begin{figure}[!t]
\centering
\includegraphics[width=3.0in]{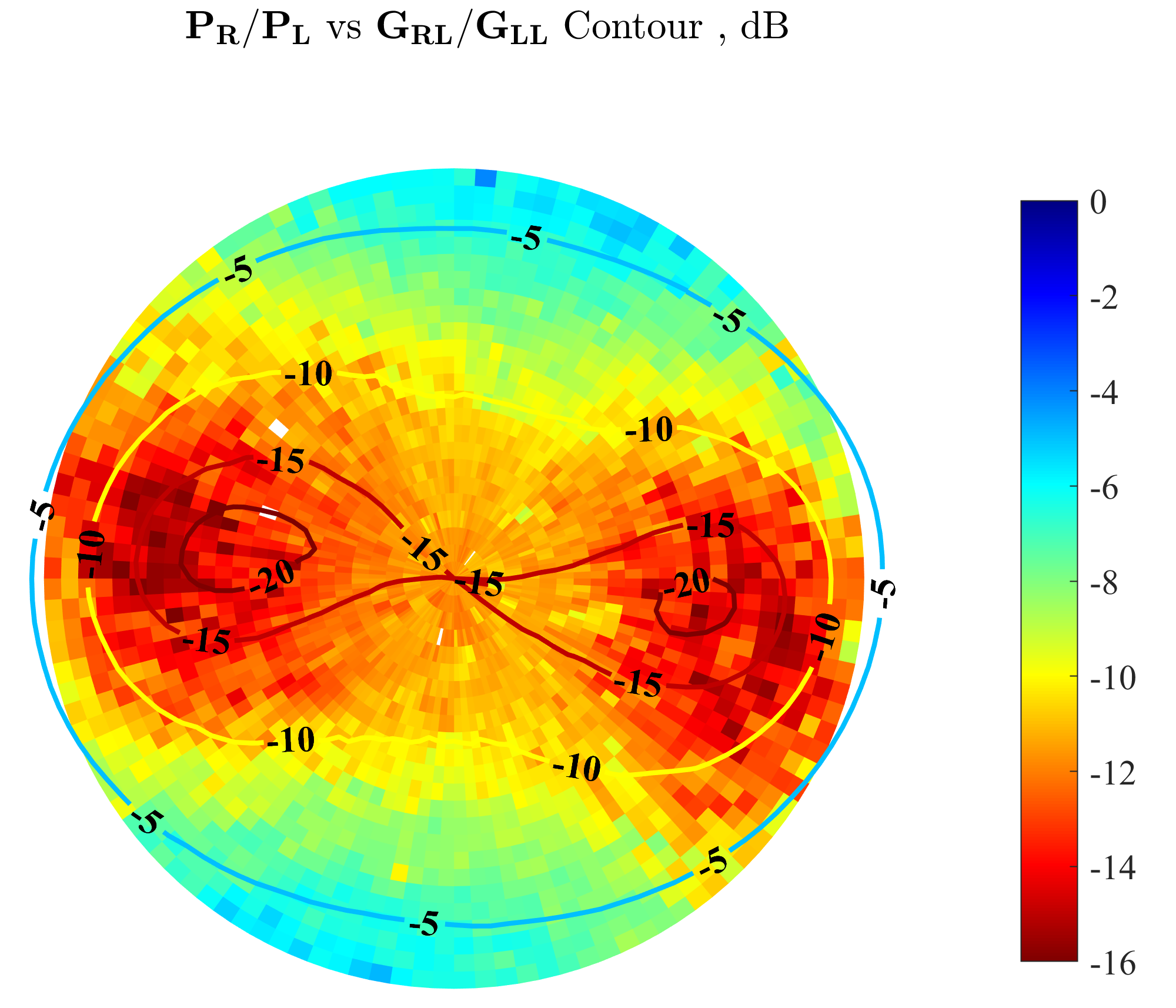}
\caption{Binned and averaged measured SP power cross-pol ratio (background pixel), plotted vs the contour of pre-launch measured antenna cross-pol ratio pattern with $48^\circ$ azimuth rotation.}
\label{pwr vs ant contour}
\end{figure}

\begin{figure}[!t]
\centering
\includegraphics[width=3.1in]{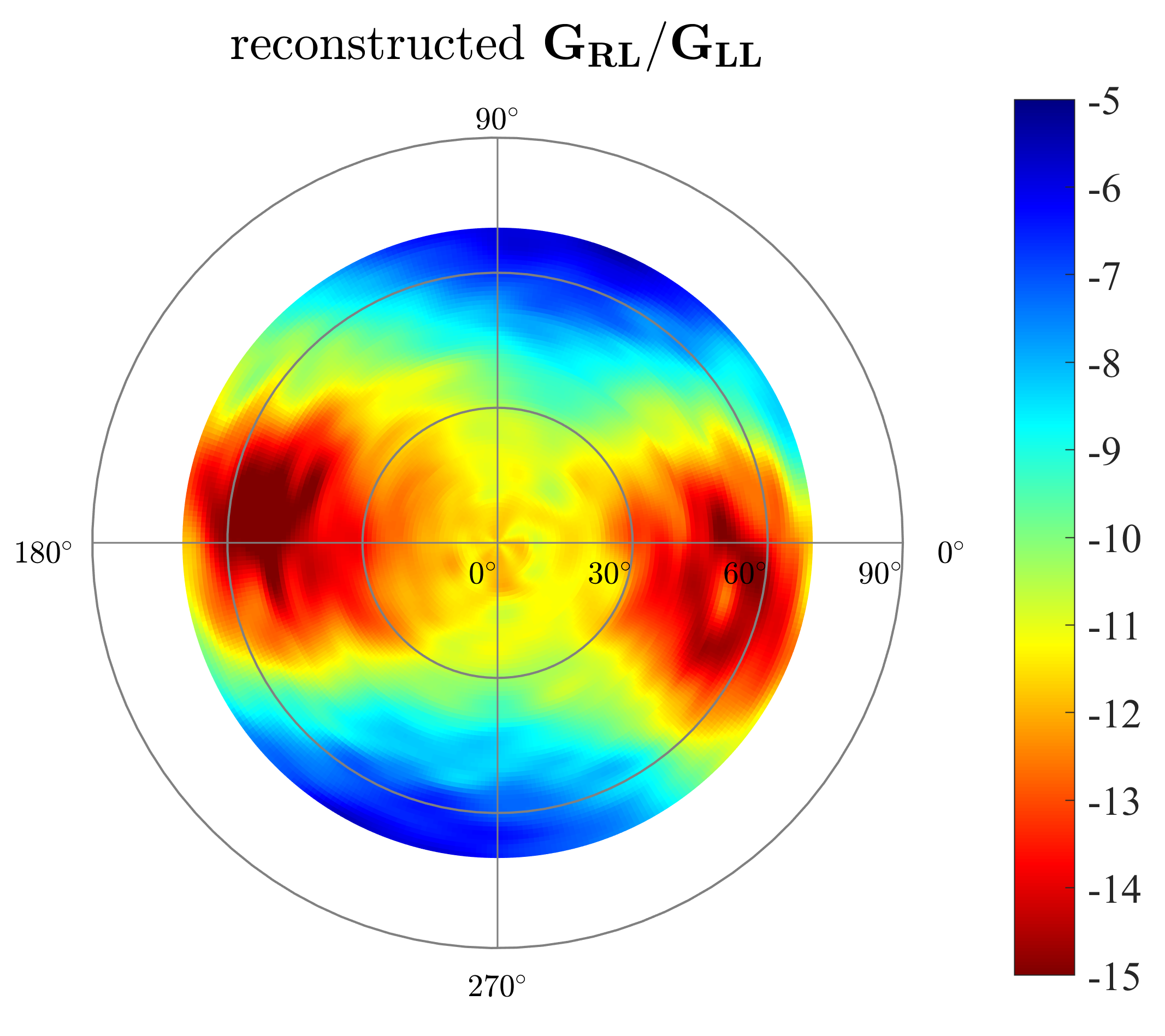}
\caption{Reconstructed antenna cross-pol ratio pattern based on onboard behaviour over ocean surface.}
\label{new xpol ratio}
\end{figure}

Comparing Fig. \ref{refl cal} (a) and (b), most of the anomalous groups of retrieved reflectivity from Block IIR are brought back below the theoretical max, with a very small amount of data higher than the theoretical max, which can be interpreted as resulting from measurement noise. However, SVN 73 still has a group of retrieved reflectivity much higher than the theoretical max, due to its inconsistent results at 30-40 degrees or 50-60 degrees incidence angle, which might result from either the uncertainty in the surface condition or the flex power mode \cite{gps_power_flex} of Block IIF GPS satellites. The Rongowai system is not able to capture the GPS power flex without calibrating its zenith channel, which was originally designed for navigation only. This problem could be mitigated by using CYGNSS calibrated dynamic EIRP data \cite{dynamic_EIRP}; this is future work. \par

The distribution of retrieved reflectivity after EIRP calibration is shown in Fig. \ref{refl cal} (d). In the same plot, the simulated effective reflectivity distribution based on (3) - (9) is generated by a Monte-Carlo simulation and displayed with the red curve. The simulation generates random distribution of incidence angles  whose Probability Density Function is generated from the distribution of real data, and random surface windspeeds at 10m reference height ($U_{10}$) with a mean of 1.71 m/s and a standard deviation of 0.3 m/s. Additionally, windspeed less than 0 is set to 0, while samples with significant wave height $H_s$ higher than 15 cm are discarded to maintain the coherent state. The 15 cm threshold was selected based on a previous study on CYGNSS coherency \cite{loria_model}. In the plot, the shapes of the measurement and simulation distributions are matched well. 

After incorporating the new adjustment into the EIRP lookup table and calibration model, the new power correction factor $K$ becomes -13.15 dB as a result of the second-order calibration. The correlation between model and measurement is increased to 0.78, and the RMSD is decreased to 0.8.

\begin{figure}[!t]
\includegraphics[width=3.3in]{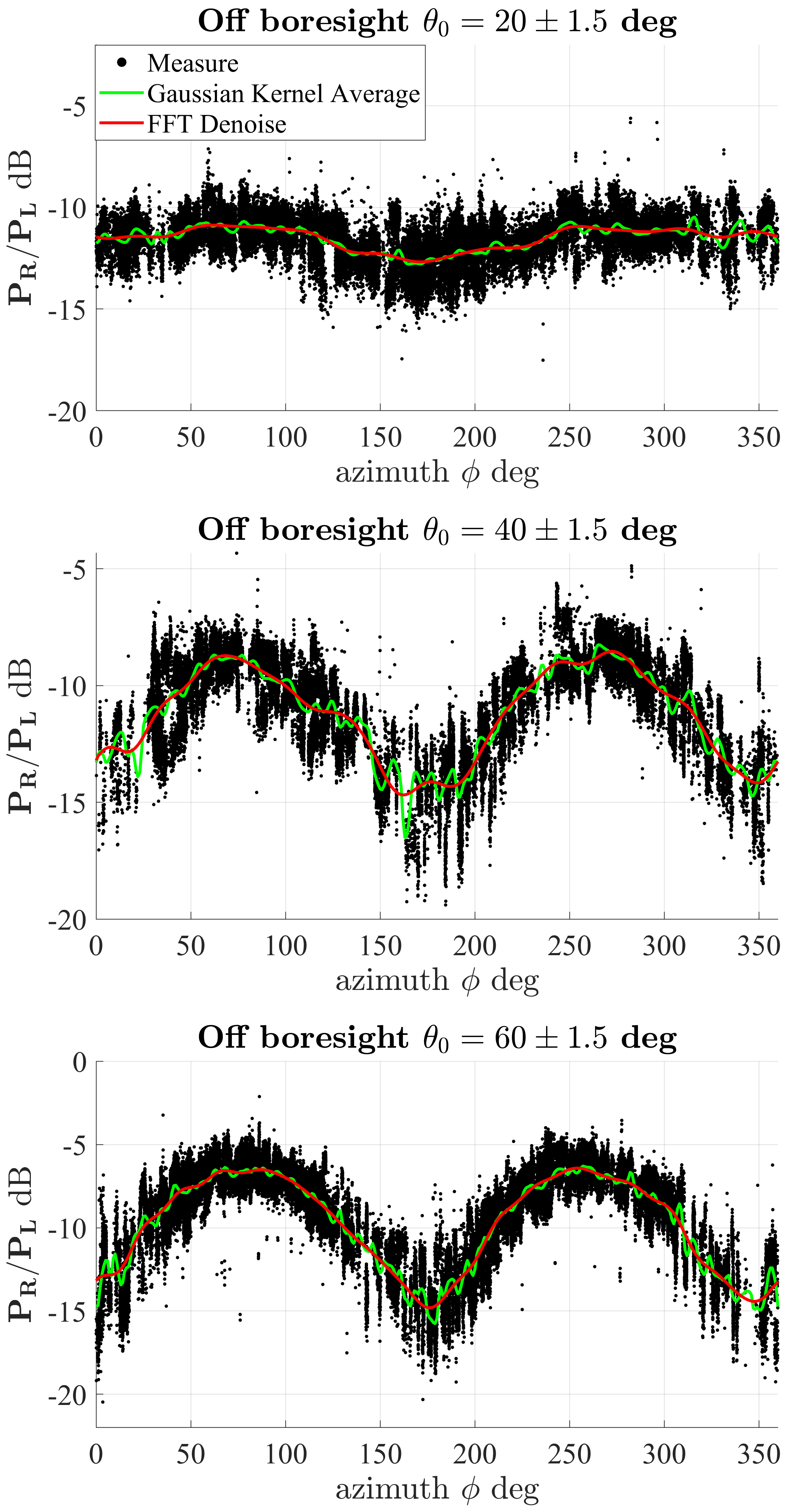}
\caption{Illustration of antenna cross-pol ratio pattern reconstruction with measured power cross-pol ratio from incoherent ocean surface. Examples of azimuth cuts at off-boresight $\theta_0$ at $20^\circ$, $40^\circ$, $60^\circ$ $\pm 1.5 ^\circ$ are shown. Measured $P_R/P_L$ (black dots) are resampled and averaged through the Gaussian kernel function (greed curve) and then denoised by FFT and lowpass filter (red curve). }
\label{ratio az cut}
\end{figure}

\subsection{Antenna Cross-pol Pattern Calibration}
\label{IIIE}
The rotation of the antenna pattern has been investigated in Section \hyperref[IIIB]{III B}. However, by comparing the measured power cross-pol ratio pattern over ocean surface and the contour of the rotated pre-launch antenna pattern as shown in Fig. \ref{pwr vs ant contour}, we notice that they differ from each other near the regions with high antenna cross-pol isolation: near the -20 dB pre-launch cross-pol pattern contour, the measured power cross-pol ratio only reaches -15 dB; while near the -10 dB pre-launch cross-pol pattern contour, the measured power cross-pol ratio becomes even lower by 3 dB. It is very likely that the antenna cross-pol pattern has been corrupted due to its coupling with the aircraft body and cannot be reliably determined using the pre-launch pattern. Therefore, we use the measured power cross-pol ratio obtained from ocean incoherent data to calibrate the antenna cross-pol ratio empirically. \par

\begin{figure*}[!b]
\centering
\includegraphics[width=\textwidth]{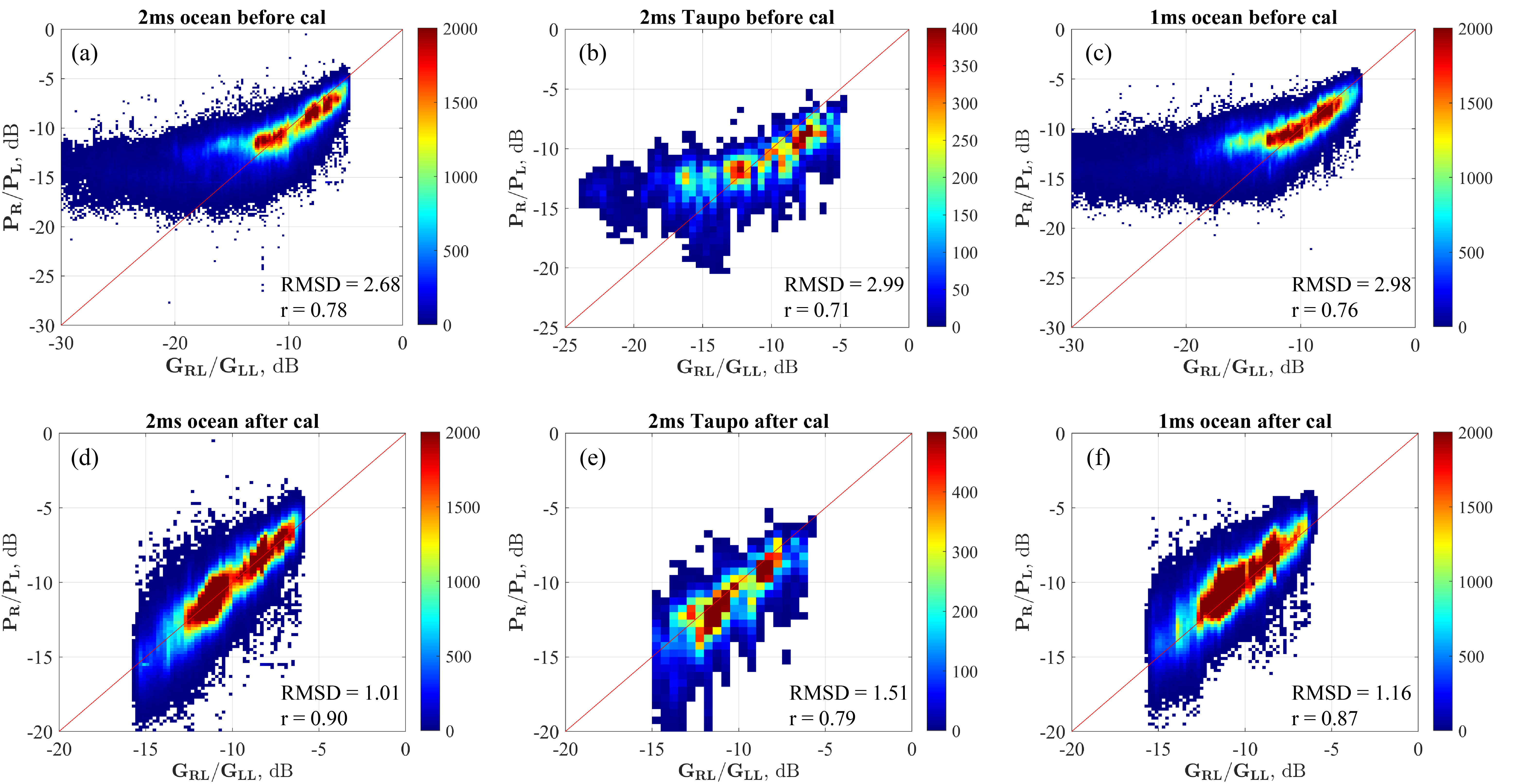}
\caption{Measured SP power cross-pol ratio $P_R/P_L$ vs antenna gain cross-pol ratio $G_{RL}/G_{LL}$ from ocean data with 2ms coherent integration time before (a) and after (d) calibration, Taupo data with 2ms coherent integration time before (b) and after (e) calibration, and ocean data with 1ms coherent integration time before (c) and after (f) calibration}
\label{xpol cal performance}
\end{figure*}

\begin{table}[!t]
\centering
\caption{Summary of parameters used in pattern smoothing}
\label{smoothing parameters} 
\begin{tabular}{|l||c||c||c|} 
\hline
\backslashbox[2.3cm]{$\theta_0$}{parameter} & \textbf{$\sigma_\theta$} & \textbf{$\sigma_\phi$} & \textbf{$f_{\mathrm{cut}}$}  \\

\hline
$\theta_0 < 20^\circ$ & 1.5 & 2 & 0.02 \\
$20^\circ < \theta_0 < 40^\circ$ & 1.5 & 1.5 & 0.03 \\
$40^\circ < \theta_0 < 50^\circ$ & 1.5 & 1.5 & 0.04\\
$50^\circ < \theta_0 < 60^\circ$ & 1.5 & 1 & 0.05 \\
\hline
\end{tabular}
\end{table}

As discussed in section \ref{IIB}, the power cross-pol ratio is the sum of the antenna cross-pol and surface Fresnel coefficient ratio (see equation (12) and (13) ). The surface Fresnel Coefficient ratio $\Gamma_{RR}/\Gamma_{LR}$   can be neglected if it is much smaller than the antenna cross-pol ratio $G_{RL}/G_{LL}$, so that the measured power cross-pol ratio $P_R/P_L$ can directly be used to calibrate the actual on-board antenna cross-pol pattern. Previous study \cite{Buchanan2019} has shown that the SMAP- R system has difficultly observing any RHCP signal above the noise floor in its RHCP DDM over ocean in both a typical sea state with approximately 7 m/s wind speed and in a very rough sea state during a hurricane with wind speed higher than 35 m/s. The SMAP-R system has a fixed observation incidence angle at $40^\circ$ and an antenna cross-pol ratio $G_{RL}/G_{LL}$ better than 25 dB within its main beam [63]. This implies that the Rongowai measured NBRCS cross-pol ratio term in  (13) can be neglected below $40^\circ$ due to Rongowai’s worse (higher) antenna cross-pol isolation $G_{RL}/G_{LL}$ and lower $\Gamma_{RR}/\Gamma_{LR}$ which decreases rapidly as it moves to near nadir incidence angles. In addition, there is no clear evidence that an incoherent RHCP signal was detected in the Rongowai DDMs over the ocean surface below $70^\circ$: For specular measurements with high power cross-pol ratio $P_R/P_L$, the RHCP DDMs are noisy with low SNR, suggesting that the observed $P_R/P_L$ is primarily a result of the antenna cross-pol leakage $G_{RL}/G_{LL}$ rather than the surface reflection property $\Gamma_{RR}/\Gamma_{LR}$. In our first order correction for the antenna cross-pol ratio, we assume that the RHCP signal from ocean observations in the RHCP channel can be ignored, and reconstruct the antenna cross-pol ratio pattern from the measured power cross-pol ratio based on (13). The procedure for this approach to determining the antenna cross-pol gain pattern is described below:

\begin{enumerate}
    \item select incoherent ocean samples with off boresight angles between $\theta_0 \pm 1.5$ degrees to form an azimuth cut of $P_R/P_L$ samples across the $2\pi$ hemisphere. The example azimuth cuts of $20 \pm 1.5^\circ$, $40 \pm 1.5^\circ$, $60 \pm 1.5^\circ$ are shown as black dots in Fig. \ref{ratio az cut}. 
    \item resample and average the $P_R/P_L$ samples in the azimuth cut with $1^\circ$ increment by a Gaussian kernel, and the averaged $P_R/P_L$ (denoted by $\Bar{G}$) in the resampled grid $(\theta_0,\phi_0)$ is calculated by:
        \begin{equation}
            \Bar{G}(\theta=\theta_0;\phi=\phi_0) = \frac{\sum_{i} \frac{P_R^i}{P_L^i} \cdot K_{\theta\phi}^i}{\sum_i K_{\theta\phi}^i}
        \end{equation}
        where
        \begin{align}
            K_{\theta\phi}^i = &\frac{1}{2\pi \sigma_{\theta} \sigma_{\phi}} \nonumber \\
            &\exp{ \left( -\frac{(\theta_i-\theta)^2 }{2\sigma_\theta^2} - \frac{(\phi_i-\phi)^2}{2\sigma_\phi^2}  \right)}
        \end{align}
    where $i$ denotes $i^{th}$ sample; $\theta$ and $\phi$ are off-boresight and azimuth angle respectively with the subscript 0 indicating the resampled grid; $\sigma_\theta$ and $\sigma_\phi$ are the the Gaussian kernel width in $\theta$ and $\phi$ dimensions respectively, and their values are summarized in table \ref{smoothing parameters}. The $2\pi$ periodicity is retained by extending the periodic data below and above its ends. The Gaussian kernel averaged results are plotted as green curves in Fig. \ref{ratio az cut}. 
    \item The Gaussian kernel averaged results are still noisy, while the antenna pattern should be smooth as installed on the aircraft with its smooth and rolled edge metal body as its ground plate. $\Bar{G}(\theta_0,\phi_0)$ is then transformed to its frequency  domain, and a low-pass filter with the cutoff frequency summarized in table \ref{smoothing parameters} is applied to exclude the high-frequency noise in its spectrum. The final smoothed results of each azimuth cut is plotted as red curve in Fig. \ref{ratio az cut}.
    \item The process is repeated for each azimuth cut with off boresight angle $\theta_0$ from 0 to 70 degrees. The reconstructed antenna cross-pol ratio pattern is shown in Fig. \ref{new xpol ratio}
\end{enumerate}

The final result of the reconstructed antenna cross-pol ratio pattern has azimuth and off-boresight resolutions of $1^\circ$. The antenna co-pol gain $G_{LL}$ is less likely to be affected by the aircraft body, so its on board antenna pattern is assumed to be the same as its pre-launch pattern other than the $48^\circ$ azimuthal rotation described in Section \hyperref[IIIB]{III B} . The reconstructed antenna cross-pol gain pattern $G_{RL}$ is obtained by multiplying the reconstructed antenna cross-pol gain $G_{RL}/G_{LL}$ by its co-pol gain $G_{LL}$. Assuming reciprocity, $G_{LR} = G_{RL}$. \par

Power measurements made at incidence angle above $65^\circ$ might be affected by multi-path propagation, and typically those samples are discarded in GNSS-R land applications. When reconstructing the antenna cross-pol pattern, samples up to $70^\circ$ are used to provide redundancy. It is still not clear whether the cross-pol reflectivity or NBRCS in (12) and (13) can be observed from ocean surfaces at high incidence angles. Therefore, the azimuth cut for each off-boresight angle is used to reconstruct the antenna pattern in order to examine the potential constant bias caused by co-pol $RR$ reflectivity or NBRCS at a fixed incidence angle (which approximately equals to the off-boresight angle) in the future second-order antenna pattern correction work. Different values of the Gaussian kernel width and cutoff frequency are adjusted for each azimuth cut with different sample number density to smooth the pattern while keeping the real cross-pol structure of the data. The filtering of the data used to reconstruct the pattern is minimal because there are more than 1 million samples obtained from ocean, large enough to ignore non-ideal samples such as those with land contamination, too low or too high winds in the averaging process of step 2. Nevertheless, the reconstructed cross-pol pattern still achieves good continuity in the off-boresight and azimuth dimensions, mainly due to the Gaussian kernel averaging. \par

Density scatter plots of the power cross-pol ratio $P_R/P_L$ versus antenna cross-pol ratio $G_{RL}/G_{LL}$ are shown in Fig. \ref{xpol cal performance}, before (a-c) versus after (d-f) calibration, with the red line as the 1:1 perfect fitting. The data with 2 ms coherent integration time obtained from ocean (a,d) is applied to reconstruct the antenna cross-pol pattern as the training set, while data with 2 ms coherent integration time obtained from Lake Taupo (b,e) and with 1 ms coherent integration time (collected between 10/1/2022 to 4/5/2023) obtained from ocean are used as the testing set to validate the performance. After calibration, all of the training and testing sets indicate a significant decrease in RMSD while an increase in Pearson correlation coefficient, which suggests that the improvement in the performance after pattern correction is consistent between different places and times. \par

\begin{figure}[!b]
\centering
\includegraphics[width=3.3in]{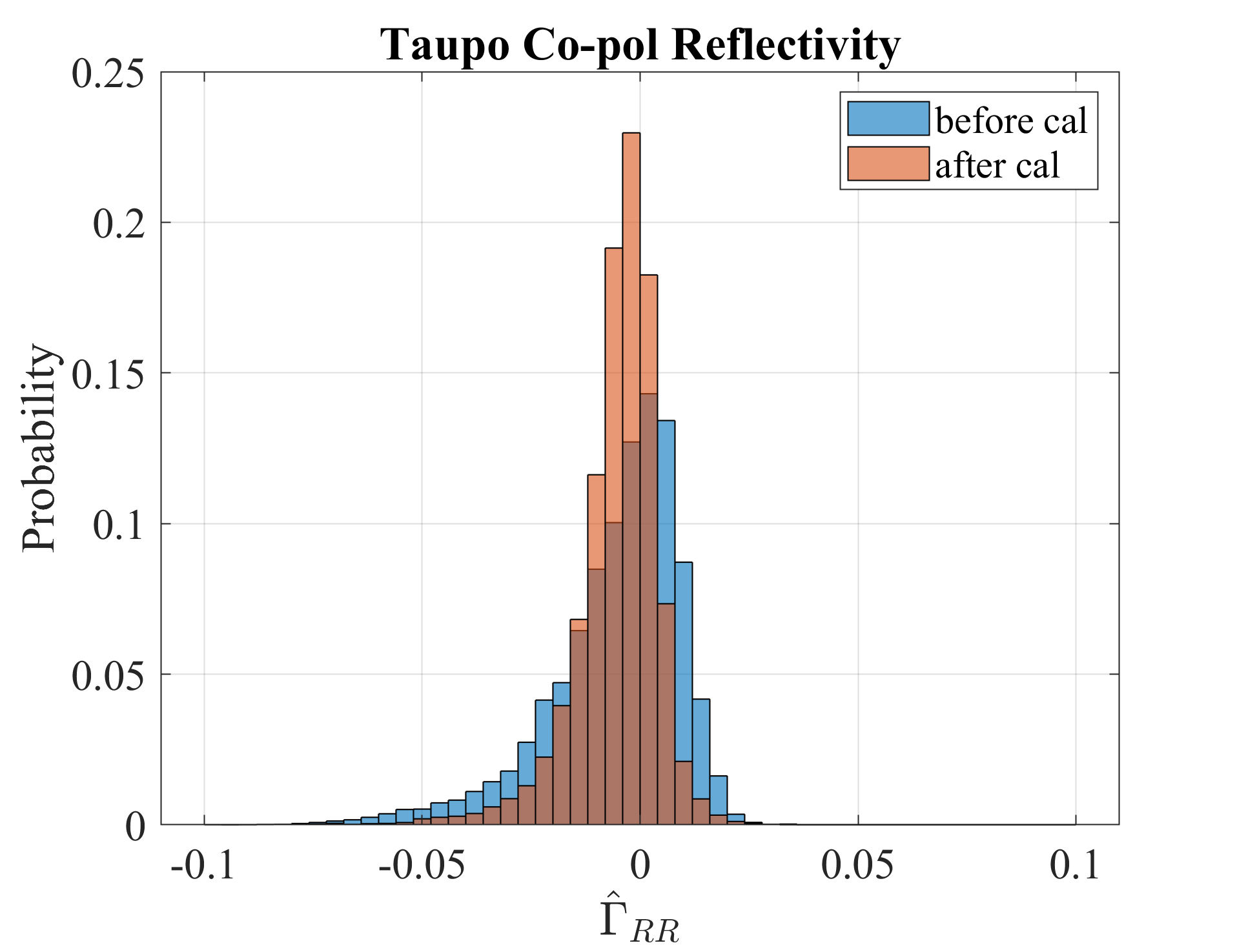}
\caption{Distributions of retrieved surface effective co-pol reflectivity from Lake Taupo measurements, before (blue) vs after (orange) calibration}
\label{Taupo xpol Refl}
\end{figure}

When both observed power cross-pol ratio $P_R/P_L$ and pre-launch antenna cross-pol pattern $G_{RL}/G_{LL}$ suggest a high cross-pol leakage (i.e. high $G_{RL}/G_{LL}$ ratio), the corrected antenna cross-pol ratio from observed $P_R/P_L$ is more trustworthy since the power in the RHCP co-pol channel is dominated by antenna LHCP-RHCP leakage. However, the improvement of the calibration performance shown in Fig. \ref{xpol cal performance} should be examined in a more careful way at low $P_R/P_L$ parts (the red pixels area in Fig. \ref{pwr vs ant contour}) due to the noise floor at the RHCP measurements:

\begin{itemize}
    \item If $P_R/P_L$ is smaller than the pre-launch cross-pol pattern $G_{RL}/G_{LL}$ at some angles (e.g. the intersected parts of the -5 dB and -10 dB contour above the red pixels in Fig. \ref{pwr vs ant contour}), we have greater confidence that the pre-launch cross-pol pattern at this part is changed to a lower value since $P_R/P_L$ is supposed to be larger than or equal to $G_{RL}/G_{LL}$ (see (13)). 
    \item  However, if $P_R/P_L$ is larger than the pre-launch cross-pol pattern $G_{RL}/G_{LL}$ (e.g. the -20 dB contour in Fig. \ref{pwr vs ant contour}), it is inconclusive that the on-board $G_{RL}/G_{LL}$ becomes larger and suffers a degraded antenna cross-pol isolation, because the lowest detectable $P_R$ is limited by the noise floor of the RHCP channel measurement. This could be the cause of the saturated $P_R/P_L$ at -15 dB with pre-launch $G_{RL}/G_{LL}$ from -15 to -30 dB shown in Fig. \ref{xpol cal performance} (a) to (c). Also the corrected $G_{RL}/G_{LL}$ has the lowest value at near -16 dB.
\end{itemize}

\begin{table}[!b]
\centering
\caption{Summarized Statistics for Different Angles and Land Types}
\label{refl summary}
\begin{tabular}{|c|c|c|c|c|c|c|c|}
\hline
\multicolumn{8}{|c|}{\textbf{$\theta_i < 35^\circ$}} \\ \hline
 &  & \textbf{Water} & \textbf{Urban} & \textbf{Crop} & \textbf{Grass} & \textbf{Shrub} & \textbf{Forest} \\ \hline
\multirow{2}{*}{$\hat{\Gamma}_{\mathrm{LR}}$} & $\mu$ & -8.81 & -13.44 & -12.15 & -15.34 & -17.64 & -17.75 \\ \cline{2-8} 
 & $\sigma$ & 3.80 & 2.86 & 2.74 & 3.47 & 3.32 & 3.47 \\ \hline
\multirow{2}{*}{$\hat{\Gamma}_{\mathrm{RR}}$} & $\mu$ & -26.67 & -25.71 & -26.00 & -25.98 & -25.42 & -25.40 \\ \cline{2-8} 
 & $\sigma$ & 4.83 & 3.86 & 4.45 & 3.58 & 2.92 & 2.99 \\ \hline
\multirow{2}{*}{$\frac{\hat{\Gamma}_{\mathrm{LR}}}{\hat{\Gamma}_{\mathrm{RR}}}$} & $\mu$ & 17.86 & 12.27 & 13.86 & 10.46 & 7.78 & 7.65 \\ \cline{2-8} 
 & $\sigma$ & 6.22 & 4.61 & 5.13 & 5.15 & 4.61 & 4.71 \\ \hline
\multicolumn{8}{|c|}{\textbf{$35^\circ < \theta_i < 50^\circ$}} \\ \hline
 &  & \textbf{Water} & \textbf{Urban} & \textbf{Crop} & \textbf{Grass} & \textbf{Shrub} & \textbf{Forest} \\ \hline
\multirow{2}{*}{$\hat{\Gamma}_{\mathrm{LR}}$} & $\mu$ & -9.09 & -14.40 & -12.72 & -15.68 & -17.84 & -17.97 \\ \cline{2-8} 
 & $\sigma$ & 4.18 & 2.77 & 2.83 & 3.44 & 3.34 & 3.41 \\ \hline
\multirow{2}{*}{$\hat{\Gamma}_{\mathrm{RR}}$} & $\mu$ & -25.70 & -24.27 & -22.89 & -24.72 & -25.34 & -25.23 \\ \cline{2-8} 
 & $\sigma$ & 4.84 & 3.56 & 4.38 & 3.84 & 3.20 & 3.22 \\ \hline
\multirow{2}{*}{$\frac{\hat{\Gamma}_{\mathrm{LR}}}{\hat{\Gamma}_{\mathrm{RR}}}$} & $\mu$ & 16.61 & 9.87 & 10.17 & 9.04 & 7.50 & 7.25 \\ \cline{2-8} 
 & $\sigma$ & 6.31 & 4.01 & 4.49 & 4.59 & 4.44 & 4.30 \\ \hline
\multicolumn{8}{|c|}{\textbf{$50^\circ<\theta_i<65^\circ$}} \\ \hline
 &  & \textbf{Water} & \textbf{Urban} & \textbf{Crop} & \textbf{Grass} & \textbf{Shrub} & \textbf{Forest} \\ \hline
\multirow{2}{*}{$\hat{\Gamma}_{\mathrm{LR}}$} & $\mu$ & -10.63 & -15.72 & -13.78 & -16.19 & -17.78 & -17.82 \\ \cline{2-8} 
 & $\sigma$ & 4.40 & 2.90 & 2.81 & 3.17 & 3.15 & 3.16 \\ \hline
\multirow{2}{*}{$\hat{\Gamma}_{\mathrm{RR}}$} & $\mu$ & -23.70 & -21.99 & -19.44 & -21.76 & -23.80 & -23.63 \\ \cline{2-8} 
 & $\sigma$ & 4.96 & 3.37 & 4.08 & 3.76 & 3.11 & 3.02 \\ \hline
\multirow{2}{*}{$\frac{\hat{\Gamma}_{\mathrm{LR}}}{\hat{\Gamma}_{\mathrm{RR}}}$} & $\mu$ & 13.07 & 6.27 & 5.66 & 5.57 & 6.02 & 5.54 \\ \cline{2-8} 
 & $\sigma$ & 6.83 & 3.86 & 4.11 & 3.97 & 4.24 & 3.39 \\ \hline
\end{tabular}
\end{table}

There are two potential ways to push the boundary of $G_{RL}/G_{LL}$. Firstly, the SNR of the RHCP channel can be raised by increasing the coherent integration time and utilizing the data collected from inland lakes under strong coherent condition. In the Rongowai mission, this can be achieved by increasing the coherent integration time up to 16 ms and collecting enough data near the -20 dB contour in Fig. \ref{pwr vs ant contour}. Alternatively, it could be done by analyzing data collected from flat desert regions where the co-pol RR reflectivity over dry sands is much higher than that over water surfaces and so is easy to model. However, no such desert is found in New Zealand, but future spaceborne polarimetric GNSS-R missions (such as HydroGNSS \cite{HydroGNSS}) could collect data from large, flat deserts in other parts of the world to calibrate antenna cross-pol gain.

\begin{figure*}[!t]
\centering
\includegraphics[width=0.95\textwidth]{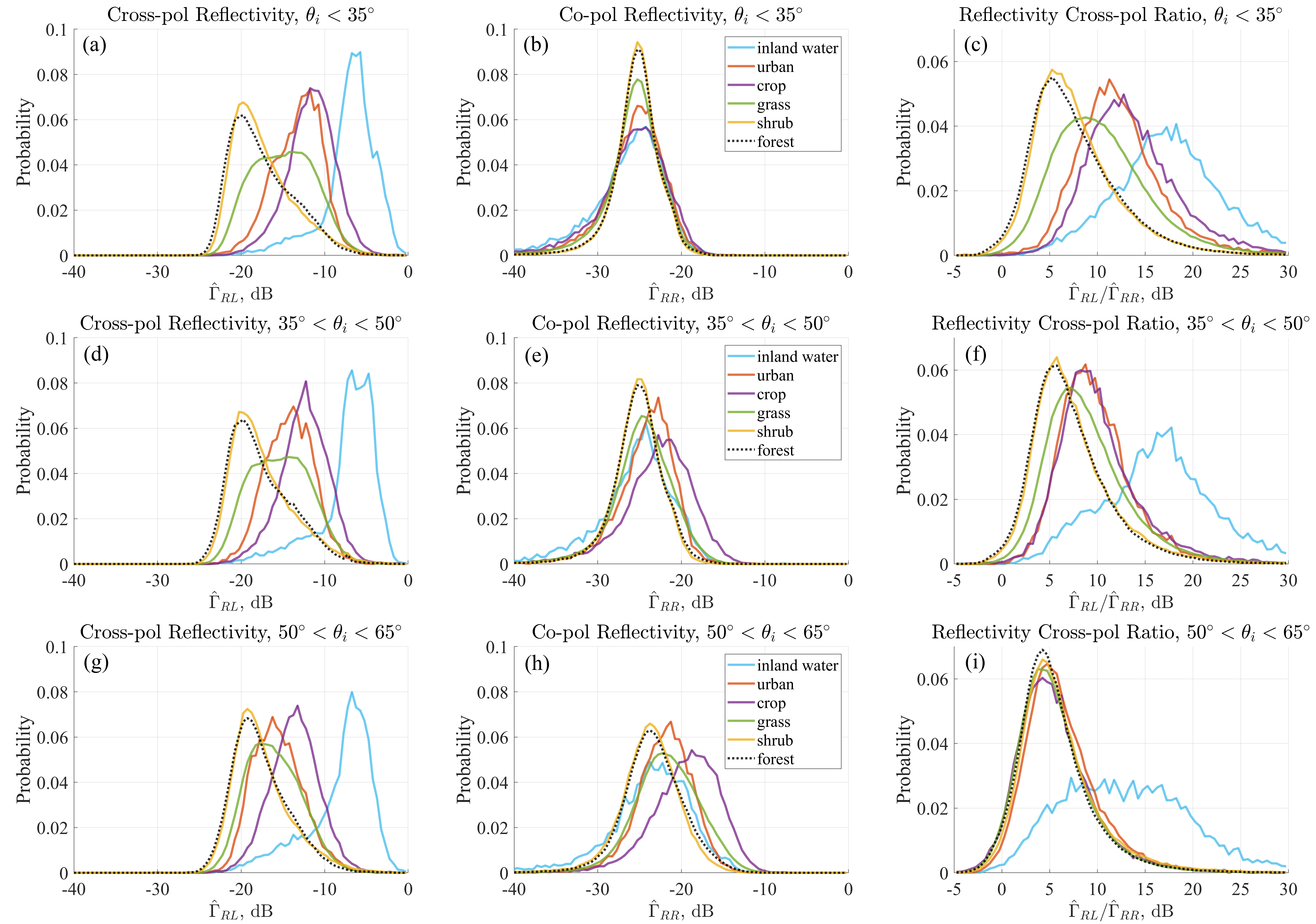}
\caption{Distributions of retrieved effective surface cross-pol (left column), co-pol (middle column) reflectivity and reflectivity cross-pol ratio (right column) of various surface types, examined with different groups of incidence angles: (a)-(c) with $\theta_i<35^\circ$; (d) to (f) with $35^\circ<\theta_i<50^\circ$; (g) to (i) with $50^\circ<\theta_i<65^\circ$}
\label{refl hist}
\end{figure*}


\section{Surface Effective Reflectivity Retrieval Result}
\label{IV}

After determining the power measurement bias, GPS transmit power, and antenna cross-pol pattern, the Rongowai L1B observables, (cross-pol and co-pol effective surface reflectivity, $\hat{\Gamma}_{LR}$ and $\hat{\Gamma}_{RR}$) can be retrieved by applying equation (14). All required calibration parameters are provided in the Rongowai L1 data files. In this section, the retrieval results of various surface types with 2 ms coherent integration time data measured over 8 months will be shown and discussed. \par

\subsection{Co-pol Reflectivity of Inland Water}
\label{IVA}
The retrieved cross-pol reflectivity $\hat{\Gamma}_{LR}$ has been examined and calibrated in Section \hyperref[IIID]{III D}. Here, the $\hat{\Gamma}_{RR}$ retrieval for an inland waterbody is examined using Lake Taupo data. All samples near shore ($<300$m) with the existing water mask are excluded to avoid land contamination. Histograms of the $\hat{\Gamma}_{RR}$ obtained from 30,000 Lake Taupo samples with different coherence states and LHCP channel SNR higher than 0 are shown in Fig. \ref{Taupo xpol Refl} in linear scale after the power bias correction and GPS transmit power calibration are applied. Different colors indicate the statistical distributions before and after antenna cross-pol pattern calibration. Both histograms look like Gaussian distributions centered at 0 because there is no cross-pol signal detected. On the other hand, in equation (14), if assuming no GPS cross-pol power (i.e. $\beta = 0$), the $\hat{\Gamma}_{RR}$ is proportional to $G_{LL}P_R - G_{RL}P_L$ which is equal to 0 if there is no RHCP-RHCP scattered signal detected. The spreading of the distribution of $\hat{\Gamma}_{RR}$ for reflections from a water surface is caused by the noise in the measurements,  and its standard deviation reflects the uncertainty in the reflectivity measurement. The standard deviation is reduced by 34\% from 0.015 to 0.01 after applying the antenna cross-pol calibration (see Fig. \ref{Taupo xpol Refl}), which  validates the effectiveness of the antenna cross-pol pattern calibration. \par

\begin{figure*}[!t]
\centering
\includegraphics[width=0.95\textwidth]{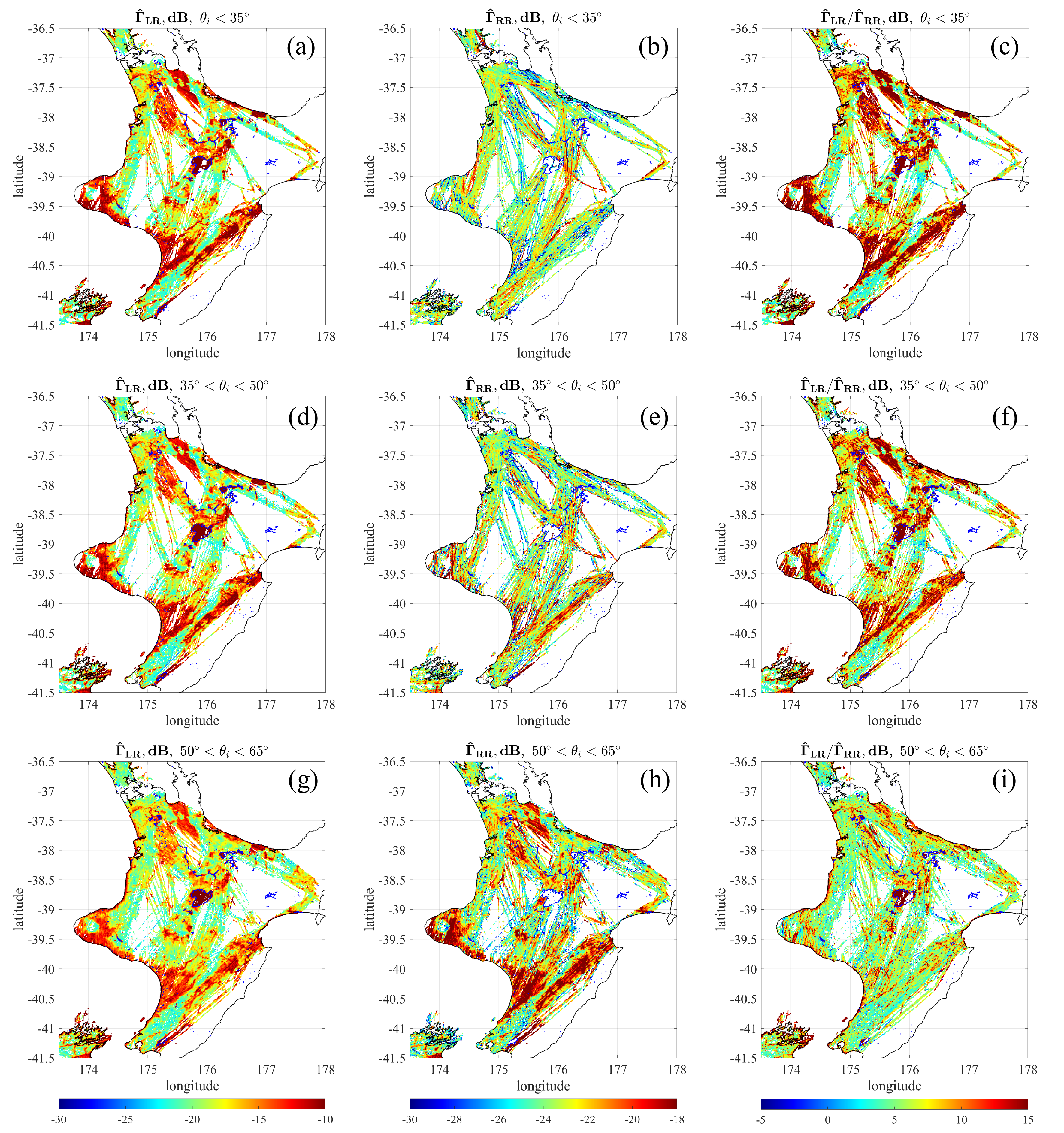}
\caption{Retrieved dual-polarized reflectivity map of NZ North Island. The retrieved cross-pol reflectivity $\hat{\Gamma}_{LR}$, co-pol reflectivity $\hat{\Gamma}_{RR}$, and reflectivity cross-pol ratio $\hat{\Gamma}_{LR}/\hat{\Gamma}_{RR}$ with different groups of incidence angles $\theta_i$ are binned and averaged within cells of 1.5 km resolution:  (a)-(c) with $\theta_i<35^\circ$; (d) to (f) with $35^\circ<\theta_i<50^\circ$; (g) to (i) with $50^\circ<\theta_i<65^\circ$}
\label{refl map}
\end{figure*}

\subsection{Surface Dual-pol Effective Reflectivity}
\label{IVB}

Each Rongowai surface measurement is assigned a specific surface type among Inland Water, Urban, Crop, Grass, Shrub, and Forest, given the land classification information at the measured specular point provided by Land Information New Zealand (LINZ) \cite{LINZ_map}. The histograms of retrieved $\hat{\Gamma}_{LR}$ and $\hat{\Gamma}_{RR}$ for each surface type across New Zealand are shown in Fig. \ref{refl hist}. These $\hat{\Gamma}_{LR}$ and $\hat{\Gamma}_{RR}$ histograms are also separated into three different ranges of incidence angle, 0 to $35^\circ$, $35^\circ$ to $50^\circ$, and $50^\circ$ to $65^\circ$ to examine the angle-dependence property of surface reflectivity. The mean values ($\mu$) and standard deviation ($\sigma$) of the distribution of each surface type and incidence angle are summarized in table \ref{refl summary}. From Fig. \ref{refl hist} (a), (d), and (g), the retrieved $\hat{\Gamma}_{LR}$ of different surface types exhibits distinct distributions: Inland water bodies have much higher $\hat{\Gamma}_{LR}$ due to the strong coherent state of the scattered GPS signal from the calm and smooth inland lakes; Shrub and Forest have distributions with similar shapes and a much lower $\hat{\Gamma}_{LR}$ than that of other surfaces. The measurements over these two types of surfaces may result from coherent scattering together with strong vegetation attenuation, or incoherent scattering, or a combination of both; $\hat{\Gamma}_{LR}$ of Grass, Crop, and Urban lie in between due to their mixed coherent and incoherent states of scattering, while the vegetation cover, soil moisture, and surface roughness are also key factors causing the variation in their retrieved effective reflectivity. Meanwhile, (a), (d), and (g) suggest that the retrieved $\hat{\Gamma}_{LR}$ of different surface types except Urban are not very sensitive to the incidence angle, which also can be reflected in the mean values summarized in table \ref{refl summary}. \par

The retrieved $\hat{\Gamma}_{{RR}}$ are shown in Fig. \ref{refl hist} (b), (e), and (h). Negative values in linear scale are excluded. Below $35^\circ$ incidence angle, all of the surface types exhibit similar distributions of $\hat{\Gamma}_{{RR}}$ as inland water, which means that there is no significant co-pol RHCP signal detected, and $\hat{\Gamma}_{{RR}}$ has a Gaussian distribution centered at 0. The distributions of the positive part of $\hat{\Gamma}_{{RR}}$ expressed in dB have the shape shown in (b). As the incidence angle increases, the $\hat{\Gamma}_{{RR}}$ distributions of different surface types shift to higher values and become more distinct from one another. This angle-dependence trend might be related to the behavior of the Fresnel coefficient. The co-pol Fresnel coefficient $\Gamma_{RR}$ is very sensitive to incidence angle, as shown in Fig. \ref{soil-water-Fresnel}. Below $40^\circ$ incidence angle,  $\Gamma_{RR}$ is much lower than  $\Gamma_{LR}$ (in the example of Fig. \ref{soil-water-Fresnel}, at $\theta_i=35^\circ$ fresh water's $\Gamma_{RR}$ is about -30 dB and soil with 10\% volumetric moisture content is about -25 dB) and drops rapidly as incidence angle decreases. Therefore, in (b) the mean values of $\hat{\Gamma}_{{RR}}$ for each surface types are considered as the minimum detectable reflectivity (MDR) and determined to be approximately -26 dB (see table \ref{refl summary}). It should be noted that the MDR sets the lower bound on detectable surface effective reflectivity, and it does not guarantee that higher reflectivity can be detected because the scattered GNSS signal available at the receiver needs reach a high enough SNR to be detectable. From (b), (e) to (h), it is observed that inland water, shrub and forest are insensitive to incidence angle below $50^\circ$. For shrub and forest, the received power may result from coherent scattering together with strong vegetation attenuation, or incoherent scattering, or a combination of both, and the extinction effect of the thick vegetation layer further reduces the SNR available at the receiver, making it harder to detect $\hat{\Gamma}_{{RR}}$ at lower incidence angles. The $\hat{\Gamma}_{{RR}}$ distribution of inland water has a larger proportion than any other surface types at the lower end. It is also observed that in (e) and (h) the higher end of the distribution of inland water $\hat{\Gamma}_{{RR}}$ increases and approaches that of urban and grass, because the observations of inland water samples include river banks and lake shores where land contamination increases the surface effective cross-pol reflectivity. Additionally, smooth surface near a water-land transition region leads to a high coherent state of the scattered signal. \par 

The retrieved effective reflectivity cross-pol ratio is shown in Fig. \ref{refl hist} (c), (f), and (i). Negative values in linear scale are excluded. The observed angle dependence results from the sensitivity of 
$\hat{\Gamma}_{RR}$ to incidence angle. Compared with other surface types, the  $\hat{\Gamma}_{LR}/\hat{\Gamma}_{RR}$ of inland water remains high while the variance increases from low to high incidence angle. $\hat{\Gamma}_{LR}/\hat{\Gamma}_{RR}$ of Grass, Urban, and Crop show significant decrease at higher incidence angles but with differing rates of decrease. As mentioned above, taking the ratio of surface effective reflectivity removes the roughness function and the vegetation extinction since they have the same effects on both cross-pol and co-pol scattering. This suggests the capability of well-calibrated $\hat{\Gamma}_{LR}/\hat{\Gamma}_{RR}$ to retrieve surface complex permittivity and inverse soil moisture independently of surface roughness and vegetation extinction.

\subsection{Retrieved Dual-pol Reflectivity Map}
\label{IVC}
Maps of calibrated $\hat{\Gamma}_{LR}$, $\hat{\Gamma}_{RR}$, and $\hat{\Gamma}_{LR}/\hat{\Gamma}_{RR}$, binned and averaged in linear scale and plotted in dB scale with 1.5 km resolution, are shown in Fig. \ref{refl map}. The North Island of New Zealand is examined in detail. Cells with negative values after averaging are discarded. The reflectivity maps are also separated into three incidence angle regions. The cross-pol reflectivity $\hat{\Gamma}_{LR}$ map is shown in the first column with some clear surface geographic features: inland water surfaces such as the Lake Taupo, Lake Rotorua, the Rangitīkei River system and the Waikato River show distinct and consistent $\hat{\Gamma}_{LR}$. Besides water, flat land surfaces such as the north, central, and southwest coastline also show high reflectivity values due to the dominant coherent state of the GPS signals scattered from these regions. In contrast, mountain areas such as volcanic Peaks (Mount Taranaki, Mount Tongariro), and axial ranges (Ruahine Range, Mount Hector) show low reflectivity since the measurements are dominated by incoherent scattering due to their rough surface features and forest coverage. \par

The co-pol reflectivity $\hat{\Gamma}_{RR}$ and reflectivity ratio $\hat{\Gamma}_{LR}/\hat{\Gamma}_{RR}$ maps are shown in the second and third columns, respectively. As shown in Fig. \ref{refl map} (d), below $35^\circ$ incidence angle, the surface $\hat{\Gamma}_{RR}$ map looks completely noise-like and does not exhibit any geophysical features since there is no significant RHCP signal detected. This agrees with the distribution shown in Fig. \ref{refl hist} (d). It is observed that as incidence angle increases, the $\hat{\Gamma}_{RR}$ rises significantly in the regions also showing high $\hat{\Gamma}_{LR}$ and dominated by coherent scattering. Due to the lower SNR in the RHCP channel, the $\hat{\Gamma}_{RR}$ map looks more noisy than the $\hat{\Gamma}_{LR}$ map. Meanwhile, the reflectivity ratio decreases as the incidence angle increases in the coherent scattering dominated regions since the $\hat{\Gamma}_{LR}$ and $\hat{\Gamma}_{RR}$ become closer to one another as the Brewster angle is approached above $60^\circ$. \par
It should be noted that in most cases the detection of RHCP scattering cannot be simply determined by inspection of the measured RHCP DDM SNR because the measurement is dominated by the receive antenna’s cross-pol leakage of the larger LHCP scattered signal. \par


\section{Discussion \& Conclusion}
\label{V}

Calibration converts raw remote sensing system measurements to meaningful physical observables with improved precision and accuracy. This can be done by comparing measurements with known targets which can be modeled accurately. For example, external targets are used to calibrate the polarimetric SAR systems, which include passive targets like trihedral corner reflectors and active sources on the ground. The general idea is to abstract the system performance into a set of calibration parameters and solve for them with the calibration model by observing the known targets. The calibration for polarimetric GNSS-R is a more difficult problem: active calibration sources cannot be used without special permission due to the restrictions on ground GPS transmitter and prevention of GPS signal jamming. Additionally, a passive calibration target is required to at least cover the first Fresnel Zone  of one GNSS-R bistatic observation as the spatial resolution in the coherent scattering mode \cite{coherent_resolution}, which is on the order of 100 m for airborne and 1km for spaceborne platforms (e.g. CYGNSS) and also depends on the coherent integration time and incidence angle. It is hard to make an artificial target to cover such a large space, and the randomness in observation angles and spatial coverage requires extra area of the target. Therefore, natural targets with large enough area and homogeneous surface materials such as ocean, inland lake, and arid desert become suitable candidates for GNSS-R calibration. Similar to the Polarimetric SAR calibration method, this paper puts emphasis on the calibration of channel cross-talk and power magnitude. Since the polarimetric GNSS-R is a type of hybrid compact polarimeter, the calibration parameters can be abstracted as a power correction factor, receiver antenna cross-pol gain, and GNSS EIRP cross-pol ratio. The power correction factor is calibrated by comparing power measurements from an inland lake with the coherent reflection model which includes the effect of water surface roughness. The receiver antenna cross-pol gain is calibrated by utilizing the measured power cross-pol ratio from the ocean surface to deduce the antenna cross-pol ratio. There is only limited information available about the GPS cross-pol EIRP level. According to GPS design specifications \cite{GPS_tech_ref} the signal Axial Rate (AR) should be no worse than 1.8 dB at L1 band. A recent publication \cite{GPS_IIF_ref} suggests that the actual AR of Block IIF GPS satellites may vary over 0.5-1.0 dB (which corresponds to $\beta$ in the range 0.08-0.30\%), while the axial ratio of other GPS block types and GNSS systems is still unclear. For the work presented here, the GPS cross-pol EIRP mix rate $\beta$ is assumed to be 0 for two reasons. The low mix rate suggested in \cite{GPS_IIF_ref} for Block IIF satellites would result in typical cross-pol leakage levels that are below our measurement noise floor, and the cross-pol level for other block types is currently unknown. If a small but non-zero GPS cross-pol EIRP were in fact present, it would have negligible impact on our calibration results for the LHCP channel at near nadir incidence angles where LHCP is the dominant scattered signal. It would, however, introduce additional uncertainty to the calibration of the RHCP channel. A proper accounting for the presence of non-zero GPS cross-pol EIRP is the subject of future work, pending the completion of ongoing ground-based measurements of GPS transmission properties for all block types.  \par

\begin{table*}[!t]
\centering
\caption{Summary of transmit power adjustment of different SVN from various incidence angles}
\begin{tabular}{|c|c|c|c|c|c|c|c|c|c|c|}
\hline
\diagbox{$\theta_i$}{\textbf{SVN}} & \textbf{41} & \textbf{43} & \textbf{44} & \textbf{45} & \textbf{48} & \textbf{50} & \textbf{51} & \textbf{53}   & \textbf{56*} \\
\hline
10$^{\circ}$-20$^{\circ}$ & -1.56 (73)  & / & /         & / & / & / & -1.31 (69) & 1.76 (1) & 3.92 (1) \\
20$^{\circ}$-30$^{\circ}$ & -1.79 (69)  & / & 0.09 (7)  & / & / & / & / & 0.57 (79) & /   \\
30$^{\circ}$-40$^{\circ}$ & /           & / & /         & / & /         & 0.10 (43) & -0.17 (26) & / & /   \\
40$^{\circ}$-50$^{\circ}$ & -2.46 (75)  & / & /         & / & 1.16 (111) & 1.15 (5) & / & 1.97 (33) & 0.95  (13) \\
50$^{\circ}$-60$^{\circ}$ & -0.69 (10)  & / & /         & 0.40 (61) & 3.39 (2) & 4.87 (3) & / & 2.91 (9) & 1.78 (8) \\
\hline
$\Delta P_T$ & -1.89 & / & / & 0.40 & 1.20 & 0.48 & -1.00 & 1.13 & /\\
\hlineB{3}
\diagbox{$\theta_i$}{\textbf{SVN}}  & \textbf{57} & \textbf{58} & \textbf{59}   & \textbf{61} & \textbf{62} & \textbf{63} & \textbf{65} & \textbf{66} & \textbf{67} \\
\hline
10$^{\circ}$-20$^{\circ}$ & / & / & -1.15 (73) & / & / & / & / & / & -0.36 (60) \\
20$^{\circ}$-30$^{\circ}$ & / & 2.94 (44) & -1.06 (86) & -1.43 (30) & / & / & / & / & /  \\
30$^{\circ}$-40$^{\circ}$ & / & 1.98 (12) & / & / & / & / & / & / & -0.29 (99)   \\
40$^{\circ}$-50$^{\circ}$ & / & 2.60 (5) & / & 0.79 (97) & 1.58 (8) & 1.05 (168) & 0.58 (3) & -0.04 (21) & -0.06 (130)\\
50$^{\circ}$-60$^{\circ}$ & 1.78 (10) & / & 0.47 (2) & 0.15 (13) & / & 2.40 (6) & 3.57 (1) & / & 1.80 (33)\\
\hline
$\Delta P_T$ & / & 2.73 & -1.08 & -1.43 & 1.58 & 1.10 & / & -0.04 & 0.00 \\
\hlineB{3}
\diagbox{$\theta_i$}{\textbf{SVN}} &\textbf{68}  &\textbf{70} & \textbf{72} & \textbf{73} & \textbf{74} & \textbf{75} & \textbf{76} & \textbf{77} & \textbf{78} \\
\hline
10$^{\circ}$-20$^{\circ}$ & / & / & / & / & / & / & / & 1.86 & 2.15  \\
20$^{\circ}$-30$^{\circ}$ & -0.42 (22) & / & 1.52 (2) & / & 1.81 (35) & / & 1.81 (8) & 0.76 (162) & 0.75 (26)  \\
30$^{\circ}$-40$^{\circ}$ & / & / & -0.49 (141) & -1.22 (18) & 1.46 (22) & 0.94 (2) & / & 1.09 (61) & /  \\
40$^{\circ}$-50$^{\circ}$ & / & /  & 0.80 (29)  & / & / & / & / & 1.43 (66) & / \\
50$^{\circ}$-60$^{\circ}$ & -0.09 (81) & 1.91 (2) & 0.04 (29) & 1.34 (25) & / & 4.76 (1) & / & -0.06 (129) & 1.33 (8) \\
\hline
$\Delta P_T$ & -0.16 & / & -0.21 & 0.27 & 1.68 & / & / & 0.68 & 1.47 \\
\hline
\end{tabular}
\label{EIRP_adj}
\end{table*}

The following methods are recommended to improve the calibration of polarimetric GNSS-R systems for future missions such as the upcoming HydroGNSS spaceborne polarimetric GNSS-R by ESA \cite{HydroGNSS}:
\begin{enumerate}
    \item The antenna's cross-pol gain might be corrupted by the platform metal body, so it is recommended to measure the pre-launch antenna pattern as it is installed on the platform or on a representative mockup. An accurately measured pre-launch pattern will provide valuable prior information when calibrating the on board pattern;
    \item The coherent reflection model estimates the scattering loss due to surface roughness caused by surface windspeed. However, only hourly recorded windspeed data on the shore were available in this study, and the true windspeed of each sample could not be estimated accurately. Secondly, the CERC model requires additional information about fetch length and water depth to accurately determine the roughness, which introduces more uncertainty. To address this problem, we recommend to distribute buoy arrays with Lidar  or ultrasonic altimeters in the target lake to estimate the surface roughness directly when the receiver passes over the region.
    \item A coherence detector based on the power DDM \cite{coherence_detector} is used to select DDMs with high coherence state for calibration. Another way is to utilize the raw IF data with phase information to separate the coherent and incoherent components of the signal, and then apply the calibration model. 
\end{enumerate}

In the SMAP-R calibration \cite{SMAP_R_calibration}, correlation and RMSD between the SMAP-R observed H/V ratio versus modeled H/V ratio with soil moisture derived by the SMAP radiometer and ERA-5 were calculated for each calibration terms to show the effectiveness of calibration. However, airborne GNSS-R observables cannot be directly compared with the SMAP or ERA-5 products due to the large difference in their measurement resolution. In-situ soil moisture monitoring sites are preferable to support the calibration/validation work of airborne GNSS-R calibration. \par

The retrieval of the L1b observables, surface effective reflectivity with LR- (cross-pol) and RR- (co-pol) polarizations, is demonstrated after calibrating the L1 dual-pol power measurement.  Retrieved cross-pol, co-pol effective reflectivity and their cross-pol ratio suggest distinct features and statistics of various surface types. The cross-pol and co-pol reflectivity map also displays clear geographic features with high resolutions of 1.5 km. Additionally, the retrieved co-pol reflectivity shows a strong incidence angle dependence. At low incidence angles below $35^\circ$, the co-pol reflectivity for all surface types can hardly be detected, and their statistics reflects the minimum detectable reflectivity and uncertainty in retrieved reflectivity. At high incidence angles above $35^\circ$, the co-pol reflectivity can be utilized with the cross-pol reflectivity together to retrieve soil moisture while separating the impacts of surface roughness and vegetation layer. Other remote sensing applications such as biomass estimation, land type classification, and cryosphere-related science applications are all possible future applications of polarimetric GNSS-R.


\appendices
\section{EIRP adjustment table}
The GPS transmit power adjustment table discussed in section \hyperref[IIID]{III D} is shown below. The number inside parenthesis indicates the number of samples used for determining the power adjustment of an SVN within a certain incidence angle window. The final $\Delta P_T$ is the weighted average over the number of samples of each incidence angle window.

\section*{Acknowledgment}

The authors would like to thank Dr. Xiaoyou Lin, Dr. Mohammad Al-Khaldi, and the Rongowai Science team for their support and contributions to the Rongowai calibration work. The authors would like to thank MetService for sharing the Taupo Aero AWS windspeed data. The authors would like to thank NASA, New Zealand’s Ministry of Business Innovation and Employment, and Air New Zealand for making the mission possible. This work was supported in part by the National Aeronautics and Space Administration Science Mission Directorate contract NNL13AQ00C with the University of Michigan.

\ifCLASSOPTIONcaptionsoff
  \newpage
\fi

\bibliographystyle{IEEEtran}
\bibliography{main.bib}

%

\begin{IEEEbiography}[{\includegraphics[width=1in,height=1.25in,clip,keepaspectratio]{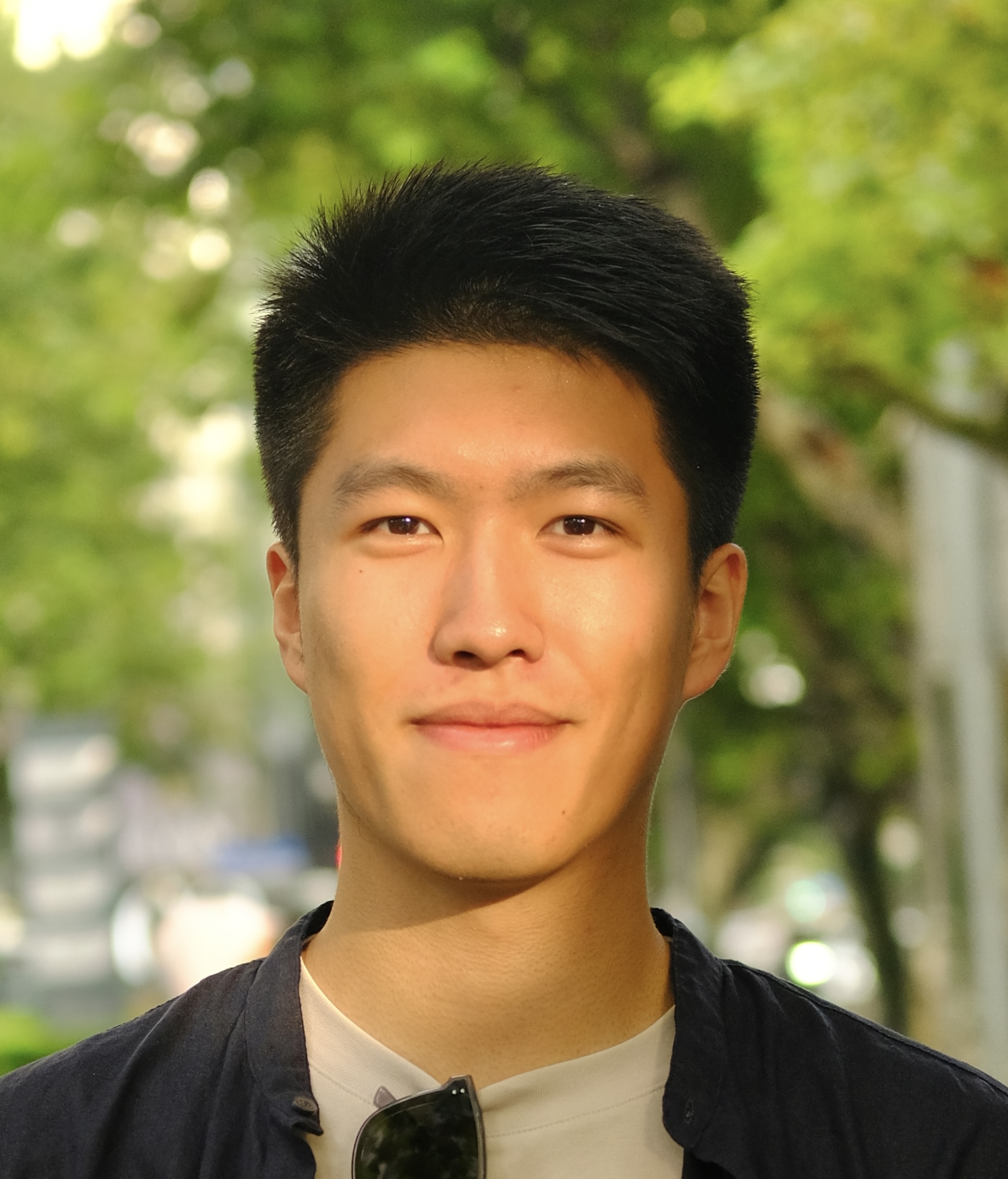}}]{Dinan Bai}
(Member, IEEE) received his B.S. degree in Electrical and Computer Engineering with Honors Research Distinction from the Ohio State University, Columbus, OH, USA, in 2022, and the M.S. degree in Electrical and Computer Engineering  from University of Michigan, Ann Arbor, USA, in 2024, where he is currently pursuing the Ph.D. degree with the Department of Climate and Space Science and Engineering. \par
Since 2022, he has been a Graduate Research Assistant with the Remote Sensing Group, Department of Climate and Space Science and Engineering, University of Michigan. His research interests include applied electromagnetics, Microwave Remote Sensing system design, calibration, and application.
\end{IEEEbiography}

\begin{IEEEbiography}[{\includegraphics[width=1in,height=1.25in,clip,keepaspectratio]{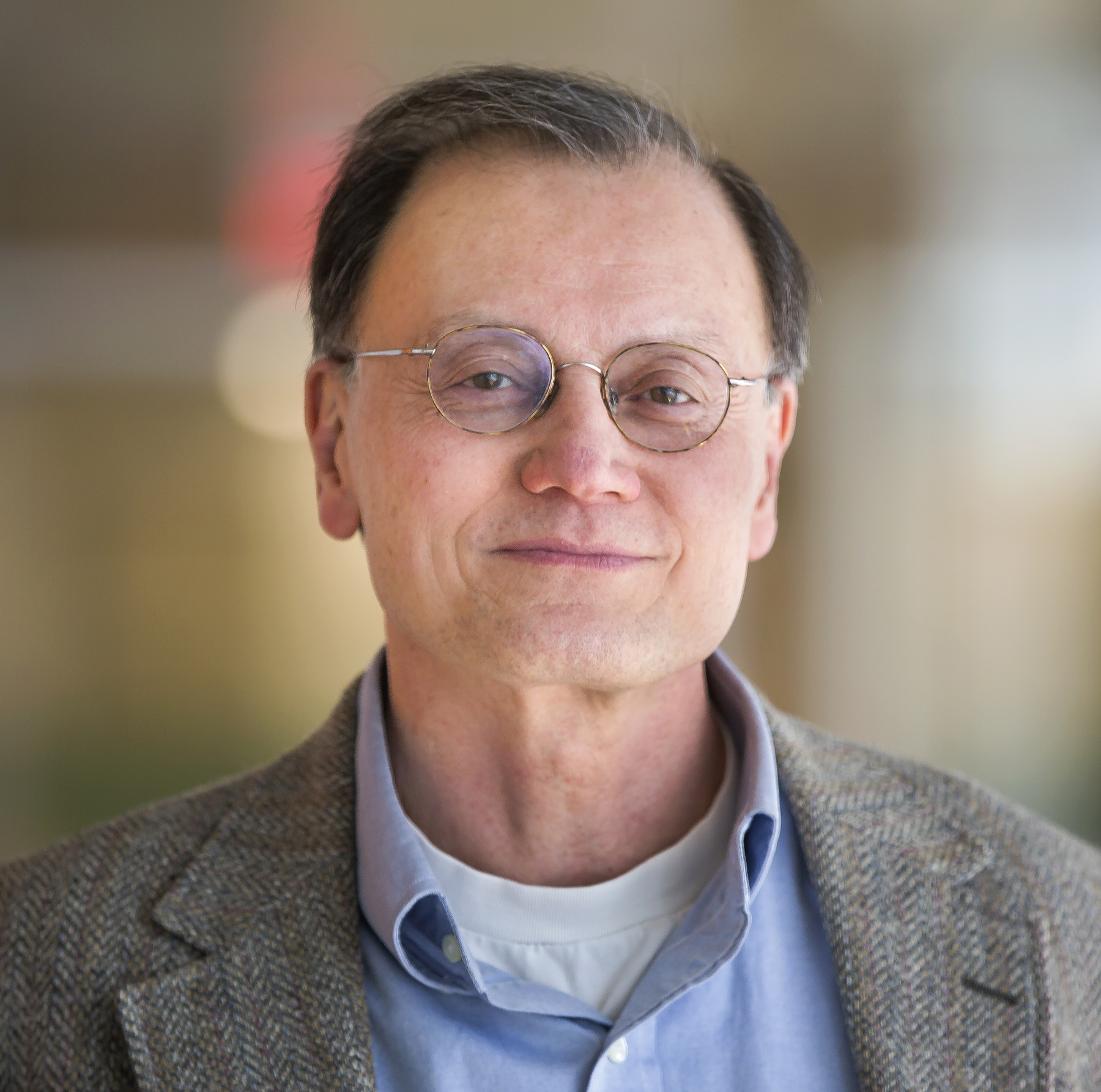}}]{Christopher S. Ruf}
(Life Fellow, IEEE) received the B.A. degree in physics from Reed College, Portland, OR, USA, in 1982, and the Ph.D. degree in electrical and computer engineering from the University of Massachusetts, Amherst, MA, USA, in 1987. \par
He is the Frederick Bartman Collegiate Professor of climate and space science with the University of Michigan and Principal Investigator of the NASA Cyclone Global Navigation Satellite System Mission. He has worked previously at Intel Corporation, Hughes Space and Communication, the NASA Jet Propulsion Laboratory, and Penn State University. His research interests include GNSS-R remote sensing, microwave radiometry, atmosphere and ocean geophysical retrieval algorithm development, and sensor technology development. \par
Dr. Ruf is a member of the American Geophysical Union, Fellow of the American Meteorological Society, and Vice Chair of Commission F of the Union Radio Scientifique Internationale. He was a recipient of the 1997 IEEE TGRS Best Paper Award, the 1999 IEEE Resnik Technical Field Award, the 2006 IGARSS Best Paper Award, the 2014 IEEE GRSS Outstanding Service Award, the 2017 AIAA SmallSat Mission of the Year Award, and the 2020 University of Michigan Distinguished Faculty Achievement Award. He is a former Editor-in-Chief of the IEEE TRANSACTIONS ON GEOSCIENCE AND REMOTE SENSING and has served on the editorial boards of Radio Science, the Journal of Atmospheric and Oceanic Technology, and Nature Scientific Reports.
\end{IEEEbiography}

\begin{IEEEbiography}{Delwyn Moller}
(Senior Member, IEEE) received the B.E. (hons) and the M.E. (Dist) degrees in Electrical and Electronic Engineering from the  University of Auckland, New Zealand and the PhD. in Electrical and Computer Engineering in 1998 from the University of Massachusetts in Amherst.   She joined NASA’s Jet Propulsion Laboratory (JPL) where she worked on radar missions and proof-of-concept initiatives.  At JPL and subsequently as a system engineer and researcher in academia, consulting and private industry, Dr. Moller has committed her career to the development of novel remote sensing systems primarily with application for Earth sciences.  She has been Principal and Co-Investigator on many research projects to include radar remote-sensing of the ocean, soil moisture, cryosphere, surface-water bodies and terrestrial surfaces .  Dr. Moller has been a member on multiple NASA mission science teams.  She is a co-recipient of the NASA Space Act award,  NASA Board Awards and multiple NASA Technology Brief awards.   \par
Dr Moller is CEO and founder of ReSTORE Lab Inc which provides development and consultancy for remote sensor development and applications with a focus on capacity building for New Zealand aerospace.  She is a Senior member of the IEEE, co-chair of the Geoscience and Remote Sensing Society (GRSS) Instrumentation and Future Technologies Technical Committee and Chair of the recently founded the New Zealand GRSS Chapter.   Through ReSTORE lab and her affiliation at the University of Auckland she is presently leading and collaborating on several international research and development efforts in advanced SAR and GNSS-R.

\end{IEEEbiography}





\end{document}